\documentclass{article}
\usepackage[a4paper, total={6in, 8in}]{geometry}
\usepackage[colorinlistoftodos]{todonotes}
\usepackage{subfig}
\usepackage{amsmath}
\usepackage{float}
\usepackage{authblk}
\usepackage{url}

\title{Complexity of products: the effect of data regularisation }

\author[1,4]{Orazio Angelini}
\author[1,2,3]{Tiziana Di Matteo}

\affil[1]{Department of Mathematics, King's College London, The Strand, London, WC2R 2LS, UK}
\affil[2]{Department of Computer Science, University College London, Gower Street, London, WC1E 6BT, UK}
\affil[3]{Complexity Science Hub Vienna, Josefstaedter Strasse 39, A 1080 Vienna, Austria}
\affil[4]{Corresponding author. Address: \texttt{angelini.orazio@gmail.com}}

\date{}

\begin{document}

\maketitle

\abstract{ Among several developments, the field of Economic Complexity (EC) has notably seen the introduction of two new techniques. One is the Bootstrapped Selective Predictability Scheme (SPSb), which can provide quantitative forecasts of the Gross Domestic Product of countries. The other, Hidden Markov Model (HMM) regularisation, denoises the datasets typically employed in the literature. We contribute to EC along three different directions. First, we prove the convergence of the SPSb algorithm to a well-known statistical learning technique known as Nadaraya-Watson Kernel regression. The latter has significantly lower time complexity, produces deterministic results, and it is interchangeable with SPSb for the purpose of making predictions. Second, we study the effects of HMM regularization on the Product Complexity and logPRODY metrics, for which a model of time evolution has been recently proposed. We find confirmation for the original interpretation of the logPRODY model as describing the change in the global market structure of products with new insights allowing a new interpretation of the Complexity measure, for which we propose a modification. Third, we explore new effects of regularisation on the data. We find that it reduces noise, and observe for the first time that it increases nestedness in the export network adjacency matrix. }


\newcommand{\logprody}[0]{logPRODY}
\newcommand{\Logprody}[0]{LogPRODY}
\newcommand{\nwkdelong}[0]{Nadaraya-Watson kernel regression}
\newcommand{\fglong}{Fitness-\gdp}
\newcommand{\clplong}{Complexity-\logprody}

\newcommand{\gdp}[0]{GDP}
\newcommand{\gdppc}[0]{GDPpc}
\newcommand{\spsb}[0]{SPSb}
\newcommand{\hmm}[0]{HMM}
\newcommand{\nwkde}[0]{NWKR}
\newcommand{\lp}[0]{L}
\newcommand{\fg}[0]{FG}
\newcommand{\clp}[0]{CL}
\newcommand{\rca}[0]{RCA}
\newcommand{\nrca}[0]{nRCA}
\newcommand{\exm}[0]{EXM}
\newcommand{\cp}[0]{_{cp}}
\newcommand{\mcp}[0]{M\cp}
\newcommand{\vfield}[0]{\vec{v}}
\newcommand{\noreg}[1]{\texttt{noreg#1}}
\newcommand{\reg}[1]{\texttt{hmm#1}}

\newcommand{\sam}[0]{s}
\newcommand{\thisc}[0]{\hat{c}}
\newcommand{\thist}[0]{\hat{t}}
\newcommand{\thisct}[0]{\thisc,\thist}
\newcommand{\analog}[0]{x}
\newcommand{\fgpos}[0]{\vec{\analog}}

\newcommand{\define}[1]{\emph{#1}} 
\newcommand{\subsubsub}[1]{\textbf{#1}. - } 
\newcommand{\qt}[1]{``#1''} 
\newcommand{\scinot}[2]{#1\times10^{#2}} 
\newcommand{\paneldeflong}[2]{\textbf{Panel #1. (#2) - }} 
\newcommand{\paneldef}[1]{\textbf{Panel #1. - }} 

\newcommand{\reffigure}[2][]{Figure#1 \ref{#2}}
\newcommand{\reffig}[2][]{Fig#1.\ref{#2}}
\newcommand{\refsection}[1]{Section \ref{#1}}
\newcommand{\refsec}[1]{Sec.\ref{#1}}
\newcommand{\refequation}[1]{Equation \ref{#1}}
\newcommand{\refeq}[2][]{Eq#1.\ref{#2}}
\newcommand{\refpanel}[1]{(#1)}
\newcommand{\reftab}[1]{Tab.\ref{#1}}
\newcommand{\reftable}[1]{Table \ref{#1}}

\section{Introduction}
\label{subsection:introduction}
Complexity and Fitness measures were originally proposed \cite{Tacchella2012} within the field of Economic Complexity (EC) to capture respectively the level of sophistication of a given class of products found on the international export market and the advancement of the productive system of a country. These two measures are calculated from international trade data, and they stem from the hypothesis that the difference between countries' competitiveness comes from their respective \emph{capabilities} \cite{Dosi1988,Lall1992,Teece1994}.
Capabilities are non-exportable features of the productive system of a country that allow it to produce a certain class of products. The problem with the theory of capabilities is that capabilities themselves are hard to define: one can speculate on what they might be, e.g. good regulations, a well-organized education system, or maybe the presence of facilities specifically useful for a product's making, but there is currently no good principled \qt{\emph{a priori}} or normative approach to classify and measure them \cite{Hidalgo2007}.
On the other hand, the observation that a country $c$ exports product $p$ contains a strong signal. It implies that $c$ is competitive enough in the production of $p$ for export to be convenient on the global market. Therefore, one could say that $c$ has all the capabilities needed to make $p$. Hausmann \cite{Hausmann2009} proposed the \define{Method of Reflections}, a non-normative algorithm to rank countries by how many capabilities they have, and products by how many capabilities they need for production, based on observed exports. The algorithm leverages topological properties of the export network, which is a bipartite network where the nodes can be either countries or product classes, and where a link is added to the network if country $c$ is a significant exporter of $p$. Fitness and Complexity are the output of an alternative algorithm \cite{Cristelli2013a} exploiting the discovery that the export network has a nested topology \cite{Tacchella2012} (a comparative analysis is found in \cite{Liao2018}). In other words, it has been observed that some countries, usually the richest in monetary terms, export almost all product classes, and some products are exported only by the countries that are most diversified in terms of export. Conversely, the less diversified countries only export a handful of products which are also being exported by almost all countries. This means that the adjacency matrix of the export network $\mcp$ can be reordered to be very close to triangular, in analogy with some biological systems \cite{Bascompte2003,Dominguez-Garcia2015}. The Fitness/Complexity algorithm takes the adjacency matrix $\mcp$ as an input and produces a value of Fitness $F$ for each country and one of Complexity $C$ for each product. Sorting the matrix rows and columns by increasing Fitness and Complexity produces the characteristic triangular structure. This ordering offers a robust way to rank the countries in terms of their competitiveness and products in terms of how sophisticated they are \cite{Tacchella2012}. Nestedness of the bipartite export network is a fundamental point of the theory and, in this paper, we measured nestedness with one widespread metric, NODF \cite{Almeida-Neto2008}, for the first time. The Economic Complexity approach is an innovative way to use the wealth of data that is being currently produced in economics, and it has the advantage of offering a data-driven and mathematically defined method of analysis, which reduces the necessity of interpretation.\\

Several results have been produced in many directions but mainly in the direction of the Fitness measure. The network approach produced an algorithm to forecast the sequence of products a country will start to export \cite{Zaccaria2014}, and inspired the exploration of innovation models \cite{Loreto2017}. Fitness as a macroeconomic indicator has been particularly fruitful. One very interesting result calls for an extension of neo-classical economic theories of growth. It is classically understood that for countries to start the process towards industrialization they have to pass a threshold of GDP per capita (\gdppc), and it has been found that higher Fitness can significantly lower this threshold \cite{Pugliese2017}.  It has long been observed that Fitness might allow for Gross Domestic Product (\define{\gdp}) prediction \cite{Tacchella2012,Cristelli2013}, but the most recent advances have introduced a dynamical systems based approach to quantitative forecasting called \define{Bootstrapped Selective Predictability Scheme} \cite{Tacchella2017} (\spsb, see \refsection{subsection:spsb}). The method is based on the observation that trajectories of countries tend to be collinear in many regions in the \gdp-Fitness two-dimensional space. Making the assumption that the growth process of countries can be modelled as a two-dimensional dynamical system allows to use nonparametric regression techniques such as the \define{method of analogues} \cite{Lorenz1969} to forecast growth. \spsb\ been proven to give state-of-the-art GDP forecasts \cite{Tacchella2018}. In this work, we prove that \spsb\ converges to a well-known nonparametric regression originally proposed by Nadaraya and Watson. The same work introduced a new regularization method for the $\mcp$ based on a Hidden Markov model (\hmm, see \refsection{subsection:hmm}), and it has been proven to give state-of-the-art \gdp\ forecasts \cite{Tacchella2018} (but, to our best knowledge, has never been applied to the Complexity measure until the present work). These ideas were originally introduced to validate the new Fitness metric, which is non-monetary, by comparing and contrasting it to an established monetary metric such as \gdp. This line of thinking proved very fruitful, so other attempts have been made to extract information by comparing an Economic Complexity metric with established ones. One such attempt compared economic inequality measurements with Fitness \cite{Sbardella2017}. 
This paper contributes to the latest developments of the Complexity and Fitness measures and it follows up mainly from the earlier work by Angelini et al. \cite{Angelini2017} focusing on the Complexity measure. In particular, the Complexity index has been paired with \define{\logprody} ($\lp$, see \refsection{subsection:logprody_def}) to obtain an interesting insight. \Logprody\ of a product is a weighted average of the \gdp\ of its exporters, where the weights are proportional to comparative advantage in making that product. It is possible to represent product classes as points on the Complexity-\logprody\ plane. Their motion on said plane can be modelled with a potential-like equation \cite{Angelini2017} (see \refsection{subsection:clpmotion} for more details). In this work, we report the results of the application of \spsb\ and \hmm\ regularization on the Complexity measure, and we show how \hmm\ affects the $\mcp$ matrices.\\

This paper is structured as follows. In \refsection{subsection:convergence} we show that, as suggested in \cite{Tacchella2017}, the \spsb\ technique converges to the faster and mathematically well-grounded \nwkdelong\ (\nwkde), allowing applications of \spsb\ to larger datasets. In \refsection{subsection:regularization_effect} we look at how the \hmm\ regularization affects the aforementioned Complexity-\logprody\ plane motion and analyse its effect on a set of different $\mcp$ matrices. Finally,  \refsection{subsection:predictions} reports our application of the \spsb\ algorithm to make predictions on the Complexity-\logprody\ plane.

\section{Results}
\subsection{Convergence of \spsb\ to a \nwkdelong}
\label{subsection:convergence}
In this section, we prove that the \spsb\ prediction method converges, for a large number of iterations, to a \nwkdelong\ (\nwkde). The idea was originally suggested in \cite{Tacchella2017}, but never developed mathematically. We prove the convergence analytically and numerically so that for all prediction purposes the two methods are interchangeable. The result is significant because it connects \spsb\ to a well-established, tried and tested technique, and frames the predictions made with this method in a more mathematically rigorous setting. \spsb\ is a non-deterministic algorithm so, at every run, it will yield slightly different results, while \nwkde\ will always produce the same results up to machine precision. From a computational perspective, \nwkde\ has much smaller time complexity, so our result allows the use of \spsb\ on much larger datasets than previously explored.\newline

\spsb\ is fundamentally a nonparametric regression. We describe the algorithm here, and in \refsection{subsection:spsb}. In the original formulation \cite{Tacchella2018}, one is presented with $\fgpos_{\thisct}$, the position of a given country $\thisc$ in the Fitness-\gdp\ (\fg) plane at time $\thist$, and wants to predict the change (displacement) in \gdp\ at the next timestep $\thist + \Delta t$, namely $\delta \analog_{\thisct}$. The method is based on the idea, advanced in \cite{Cristelli2013}, that the growth process of countries is well modeled by a low-dimensional dynamical systems. For many important cases, the best model is argued to be embedded in the two-dimensional Euclidean space given by Fitness and \gdppc. It is not possible to identify the analytical equations of motion, so instead one uses observations of previous positions and displacements of other countries $(\delta \analog_{c,t}, \fgpos_{c,t})$, which are called \define{analogues}, a term borrowed from \cite{Lorenz1969}. Because the evolution is argued to be dependent only on two parameters, observed past evolutions of countries nearby $\fgpos_{\thisct}$ in the \fg\ plane are deemed to be good predictors of $\delta \analog_{\thisct}$. Threfore \spsb\ predicts $\delta \analog_{\thisct}$ as a weighted average of past observations. The weights will be proportional to the similarity of country $\thisc$ to its analogues, and the similarity is evaluated by calculating Euclidean distance on the Fitness-\gdp\ plane. A close relative of this approach is the well-known K-nearest neighbours regression \cite{friedman2001elements}. In order to obtain this weighted average, one samples with repetition a number $B$ of bootstraps from all $N$ available analogues. The sample probability density of an analogue $\delta \analog_{c,t}$, found at position $\fgpos_{c,t}$ is given by a gaussian distribution:
\begin{align}
&p(\delta \analog_{c,t} | \fgpos_{c,t}) = \mathcal{N}(\fgpos_{\thisct} - \fgpos_{c,t} | 0, \sigma),\\
&\mathcal{N}(\vec{z}| \vec{\mu}, \sigma) = \frac{1}{\sigma \sqrt{2\pi}}\text{exp}\left(\frac{(\vec{z} - \vec{\mu})^2}{2\sigma^2}\right).
\end{align}
Therefore sampling probability will be inversely proportional to distance, i.e. analogues closer on the \fg\ plane are sampled more often. We will adopt the following notation: each bootstrap will be numbered with $b$ and each sampled analogue in a bootstrap with $n$, so each specific analogue sampled during the prediction of $\delta \analog_{\thisct}$ can be indexed with $\sam_{b,n}^{\thisct}$. Once the sampling operation is done, one averages the samples per bootstrap, obtaining $v_b^{\thisct} = \sum^N_n \sam_{b,n}^{\thisct}/N = \langle\sam_{b,n}^{\thisct}\rangle_n$. These averaged values constitute the distribution we expect for $\delta \analog_{\thisct}$. From this distribution we can derive an expectation value and a standard deviation (interpreted as expected prediction error) for $\delta \analog_{\thisct}$:
\begin{align}
&E_{\text{\spsb}}(\delta \analog_{\thisct})=\frac{1}{B}\sum^B_{b=1} v_b^{\thisct},\\
&\sigma^2_{\text{\spsb}}(\delta \analog_{\thisct})=\frac{1}{B-1}\sum_{b=1}^B \left( v_b^{\thisct} - E_{\text{\spsb}}(\delta \analog_{\thisct}) \right)^2 \\.
\end{align}Because closer analogues are sampled more, they will have a bigger weight in the averaging operations needed to compute expected value and standard deviation. The technique can be easily extended to other types of prediction, as we did in \refsection{subsection:predictions}.\newline

\nwkde\ is conceptually very similar to \spsb. We will use the symbol $\leftrightarrow$ to establish a correspondence between the two algorithms: in \nwkde\ one is presented with an observation $X\leftrightarrow\fgpos_{\thisct}$ and wants to predict $Y\leftrightarrow\delta \analog_{\thisct}$ from it. Other observations are available $(Y_i,X_i)\leftrightarrow(\delta \analog_{c,t}, \fgpos_{c,t})$, and the prediction is a weighted average of the $Y_i$'s.
\begin{equation}
 E(Y | X)=\frac{\sum_i K_h (X-X_i) Y_i}{\sum_i K_h (X-X_i)}
\end{equation}
The weights will be given by $K$, a function of the distance on the Euclidean space containing the $X_i$ values. This function is called \define{kernel}. A more detailed explanation of this technique can be found in \refsection{subsection:nwkde}.  

\subsubsection{Analytical convergence}
\spsb\ returns both an expected value and a standard deviation for the quantity being measured. We begin by proving convergence of expected value.\\

\subsubsub{Expected values}
Suppose that we execute $B$ bootstraps of $N$ samples from all available analogues $\{\delta \analog_{c,t}\}$, so that each sampled value in a bootstrap can be labelled as $\sam_{b,n}^{\thisct}$ with $1\leq n\leq N$ and $1\leq b\leq B$. Then the \spsb\ probabilistic forecast $E_{\text{\spsb}}(\delta \analog_{\thisct})$ will be:
\begin{equation}
E_{\text{\spsb}}(\delta \analog_{\thisct}) = \frac{1}{B}\sum_{b=1}^B \left( \frac{1}{N}\sum_{n=1}^N \sam_{b,n}^{\thisct} \right) = \frac{1}{BN}\sum_{b=1}^B \sum_{n=1}^N \sam_{b,n}^{\thisct}.
\label{convergence:expvalue_spsb}
\end{equation}
If we aggregate all $B$ bootstraps, we can label the frequency with which the analogue $\delta \analog_{c,t}$ appears overall in the sampled analogues as
\begin{equation}
 \phi_{B}^{\thisct}(\delta\analog_{c,t})=\frac{1}{BN}\sum_{b=1}^B \sum_{n=1}^N \mathbf{1}_{\{ \delta\analog_{c,t}=\sam_{b,n}^{\thisct}\}}
\end{equation}
where $\mathbf{1}_{\{ \cdot \}}$ is intended to be an indicator function. So we can rewrite the forecast as:
\begin{equation}
E_{\text{\spsb}}(\delta x_{\thisct}) = \sum_{c,t} \phi_{B}^{\thisct}(\delta\analog_{c,t}) \delta \analog_{c,t},
\label{eq:kernel_prob_phi}
\end{equation}
where $\sum_{c,t}$ indicates a sum over all available analogues. But since the analogues are being sampled according to a known probability distribution $p(\delta \analog_{c,t}| \fgpos_{\thisct})$, we can expect, by the law of large numbers, that for $B\rightarrow \infty$ the sample frequency will converge to the probability values (which it does, see \reffig{fig:convergence:kernel_prob}\refpanel{a}):
\begin{equation}
\phi_{B}^{\thisct}(\delta\analog_{c,t}) \xrightarrow{B\rightarrow \infty} p(\delta \analog_{c,t}| \fgpos_{\thisct})
\label{eq:phi_large_numbers}
\end{equation}
Now, \spsb\ uses a Gaussian probability distribution $p(\delta \analog_{c,t}| \fgpos_{\thisct}) = \mathcal{N}(\fgpos_{c,t}- \fgpos_{\thisct}|0,\sigma)$ (see \refsection{subsection:spsb}) so our forecast will tend to:
\begin{equation}
E_{\text{\spsb}}(\delta x_{\thisct}) \xrightarrow{B\rightarrow \infty} \sum_{c,t} p(\delta \analog_{c,t}| \fgpos_{\thisct}) \delta x_{c,t} = \sum_{c,t} \mathcal{N}(\fgpos_{c,t}- \fgpos_{\thisct}|0,\sigma) \delta \analog_{c,t}\equiv E_{\text{\nwkde}}(\delta x_{\thisct}),
\label{eq:e_nwkr_def}
\end{equation}
but this is exactly the definition of a \nwkde\ with Gaussian\footnote{Note that in the machine learning literature it's usually not called Gaussian, but \emph{radial basis function}.} kernel that has bandwidth $\sigma$ (see \refsection{subsection:nwkde}). We assumed for brevity that the sum is already normalized, i.e. $\sum_{c,t} p(\delta \analog_{c,t}| \fgpos_{\thisct}) = \sum_{c,t} \mathcal{N}(\fgpos_{c,t}- \fgpos_{\thisct}|0,\sigma) = 1$, normalization is needed in \refeq[ns]{eq:phi_large_numbers},\ref{eq:e_nwkr_def} if this is not true, but it doesn't change the result of the proof.\newline

\subsubsub{Variances}
The variance of the distribution of samples in \spsb\ is calculated first by computing $v_b^{\thisct} = \sum^N_n \sam_{b,n}^{\thisct}/N = \langle\sam_{b,n}^{\thisct}\rangle_n$ i.e. the average of the samples of each boostrap, and then computing the variance of the $v_b^{\thisct}$ across bootstraps, so (with the same notation as \refeq{convergence:expvalue_spsb}) it can be written as:
\begin{equation}
\begin{split}
\sigma^2_{\text{\spsb}} &= \frac{1}{B-1}\sum_{b=1}^B \left( \frac{1}{N} \sum_n^N \sam_{b,n}^{\thisct} - \frac{1}{BN} \sum_{b',n'}^{B,N} \sam_{b',n'}^{\thisct} \right)^2 \\
 &= \frac{1}{B-1}\sum_{b=1}^B \left( v_b^{\thisct} - E_{\text{\spsb}}(\delta x_{\thisct}) \right)^2 \\
 &\approx \frac{1}{N} \sigma^2_{bn}(\sam_{b,n}^{\thisct})\\
 &\equiv  \frac{1}{N}\left(\frac{1}{(BN-1)} \sum_{b}^{B}\sum_{n}^{N} (\sam^{\thisct}_{b,n} - E_{\text{\spsb}}(\delta x_{\thisct}))^2\right)\\
 &\approx \frac{1}{N} \left( \sum_{b}^{B}\sum_{n}^{N} \frac{(\sam^{\thisct}_{b,n})^2}{BN} - E_{\text{\spsb}}(\delta x_{\thisct})^2 \right).
\end{split}
\label{eq:nwkde_sigma_proof}
\end{equation}
In the second row we considered that $\frac{1}{BN} \sum_{b',n'}^{B,N} \sam_{b,n}^{\thisct}$, the operation of averaging across all sample analogues, irrespective of which bootstrap they are in, is equivalent to taking the expected value in \spsb. In the third row, because in \spsb\ we are calculating the variance of the means $\langle\sam_{b,n}^{\thisct}\rangle_n$, and each of the means is done over $N$ samples, for the central limit theorem when $N\gg1$ we expect a variance that is $N$ times smaller than the population variance of the analogues sampled with probability $p$, which we called $\sigma^2_{bn}(\sam_{b,n}^{\thisct})$. The approximation in the last row is justified by the fact that $\sigma^2_{bn}(\sam_{b,n}^{\thisct})$ in the third and fourth row is an unbiased estimator of the variance, and $\sum_{b,n}^{B,n}(\sam^{\thisct}_{b,n})^2/(BN)$ in the last row is an unbiased estimator of the second moment of the distribution of the samples. In the limit of large $B$, the relation $E((z-E(z))^2)=E(z^2)-E(z)^2$ applies to unbiased estimators too.\\

Now, we know by the definition of \nwkde\ (\refsection{subsection:nwkde}) that $E(\delta \analog_{\thisct}) \leftrightarrow E(Y)$ is actually a conditional probability $E(\delta \analog_{\thisct} | x_{\thisct}) \leftrightarrow E(Y|X)$, i.e. the probability of observing a certain displacement $\delta \analog_{\thisct}$ given the position on the plane $\fgpos_{\thisct}$. Therefore we can compute the variance for a \nwkde\ as:
\begin{equation}
\sigma^2(Y|X) = E(Y^2|X) - E(Y|X)^2
\end{equation}
which tranlsates, for \spsb\ formalism, into:
\begin{equation}
\begin{split}
\sigma^2_{\text{\spsb}} &= \frac{1}{N} \sigma^2_{bn}(\sam_{b,n}) \\
&\xrightarrow{B\rightarrow \infty}  \frac{1}{N} \left(\sum_{c,t} p(\delta \analog_{c,t} | \fgpos_{\thisct}) (\delta \analog_{c,t})^2 - E_{\text{\nwkde}}(\delta \analog_{\thisct})^2 \right) \\
&= \frac{1}{N} \left(\sum_{c,t} \mathcal{N}(\fgpos_{c,t} - \fgpos_{\thisct}| 0, \sigma) (\delta \analog_{c,t})^2 - E_{\text{\nwkde}}(\delta \analog_{\thisct})^2 \right) \\
&\equiv \frac{1}{N} \sigma^2_{\text{\nwkde}}.
\end{split}
\end{equation}
We again omitted normalization terms in the third and fourth rows. This equation, combined with \refeq{eq:nwkde_sigma_proof}, means that the standard deviation calculated with \nwkde\ is espected to be proportional to the standard deviation calculated with \spsb\ multiplied by $\sqrt{N}$. Note that this method makes it possible to estimate any moment of the $\hat{f}(X|Y)$ distribution, not just the second.

\subsubsection{Numerical convergence}
\begin{figure}[H]
    \centering
    \subfloat[]{
        \includegraphics[width=.49\textwidth]{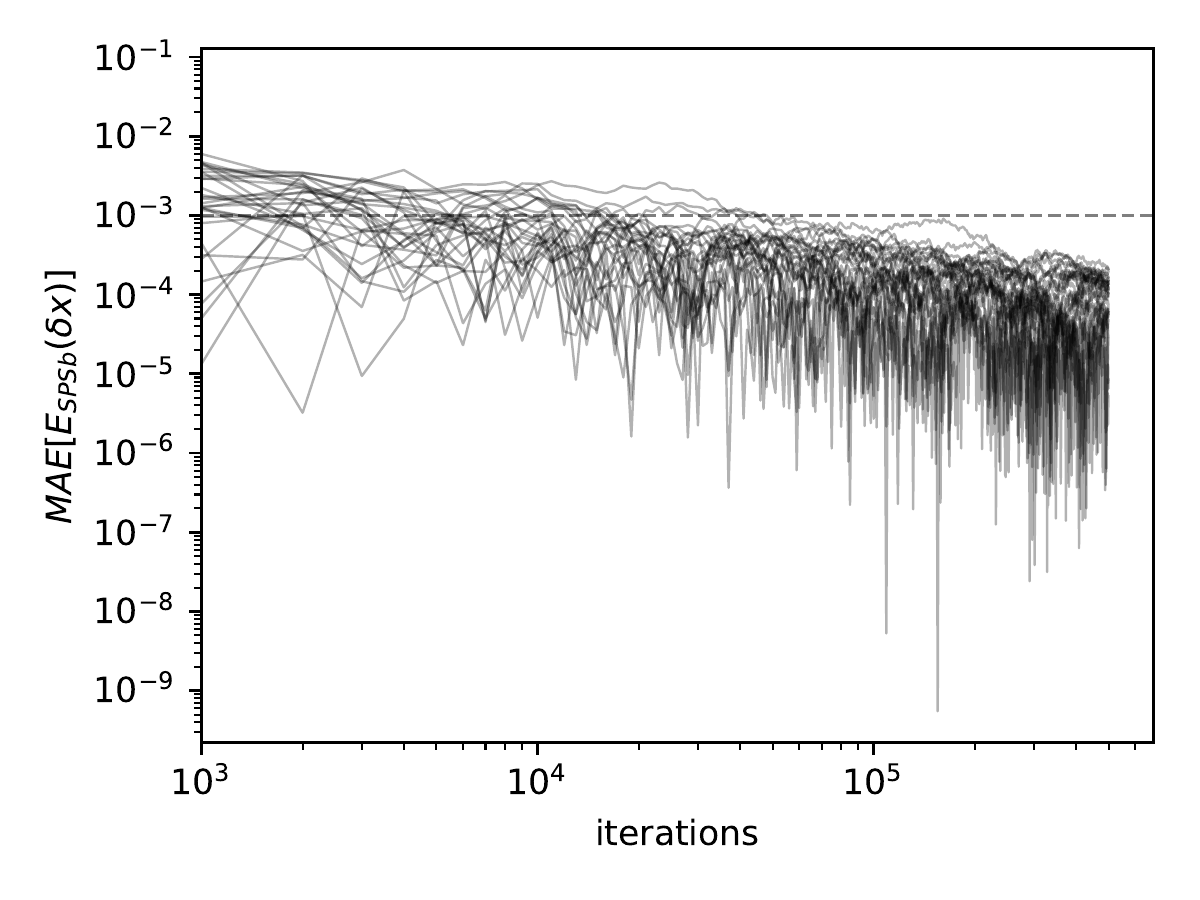}}
    \subfloat[]{
        \includegraphics[width=.49\textwidth]{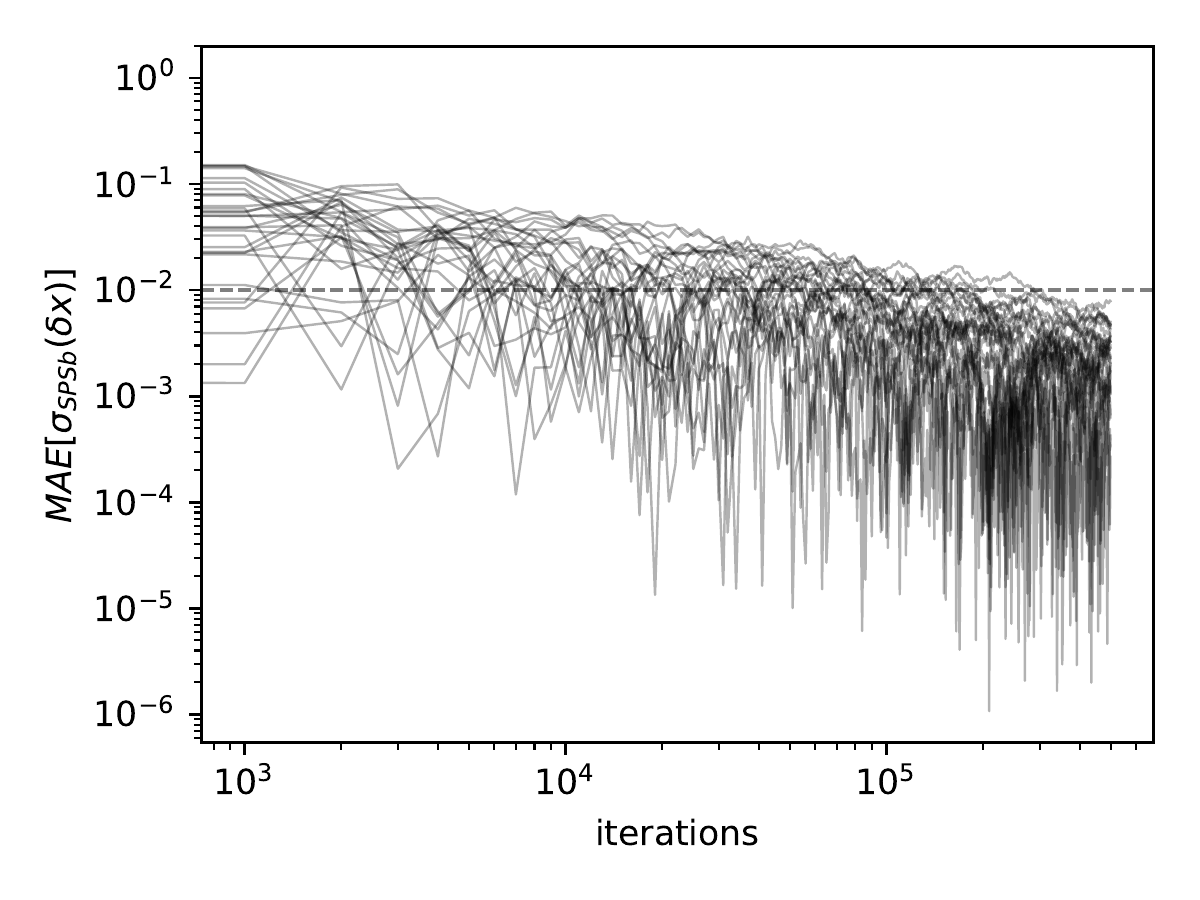}}
    \caption{\paneldeflong{a}{left} For 30 predictions, we show the difference between expectation values calculated with \spsb\ and the same quantity computed with \nwkde\ at different numbers of bootstraps. On the vertical axis, $\text{MAE}[E_{\text{\spsb}}(\delta x)] = \text{abs} \left[ \frac{E_{\text{\spsb}}(\delta x)-E_{\text{\nwkde}}(\delta x)}{E_{\text{\nwkde}}(\delta x)} \right]$, i.e. the percentage mean average error done by \nwkde\ while estimating the output of \spsb, while on the horizontal axis the number of bootstraps. After $B=10^5$ bootstrap cycles (with the default $N=100$ samples per cycle), the relative error is always smaller than 0.1\%. This figure also allows to estimate by how much \spsb\ results can vary between different runs. For $10^{3}$ bootstrap cycles, the largest deviation is around 1\% of the value.\\
    \paneldeflong{b}{right} For 30 predictions, we show the difference between standard deviations calculated with \spsb\ and the same quantity computed with \nwkde\ at different numbers of bootstrap cycles. On the vertical axis $\text{MAE}[\sigma_{\text{\spsb}}(\delta x)] = \text{abs} \left[ \frac{\sigma_{\text{\spsb}}(\delta x)-\sigma_{\text{\nwkde}}(\delta x)}{\sigma_{\text{\nwkde}}(\delta x)} \right]$, i.e. the percentage mean average error done by \nwkde\ while estimating the standard deviation predicted by \spsb, while on the horizontal axis the number of bootstraps. After $10^{5}$ bootstrap cycles, the relative error is always less than 1\%.}
    \label{fig:convergence:convergence_main}
\end{figure}

\begin{figure}[H]
    \centering
    \subfloat[]{
        \includegraphics[width=.49\textwidth]{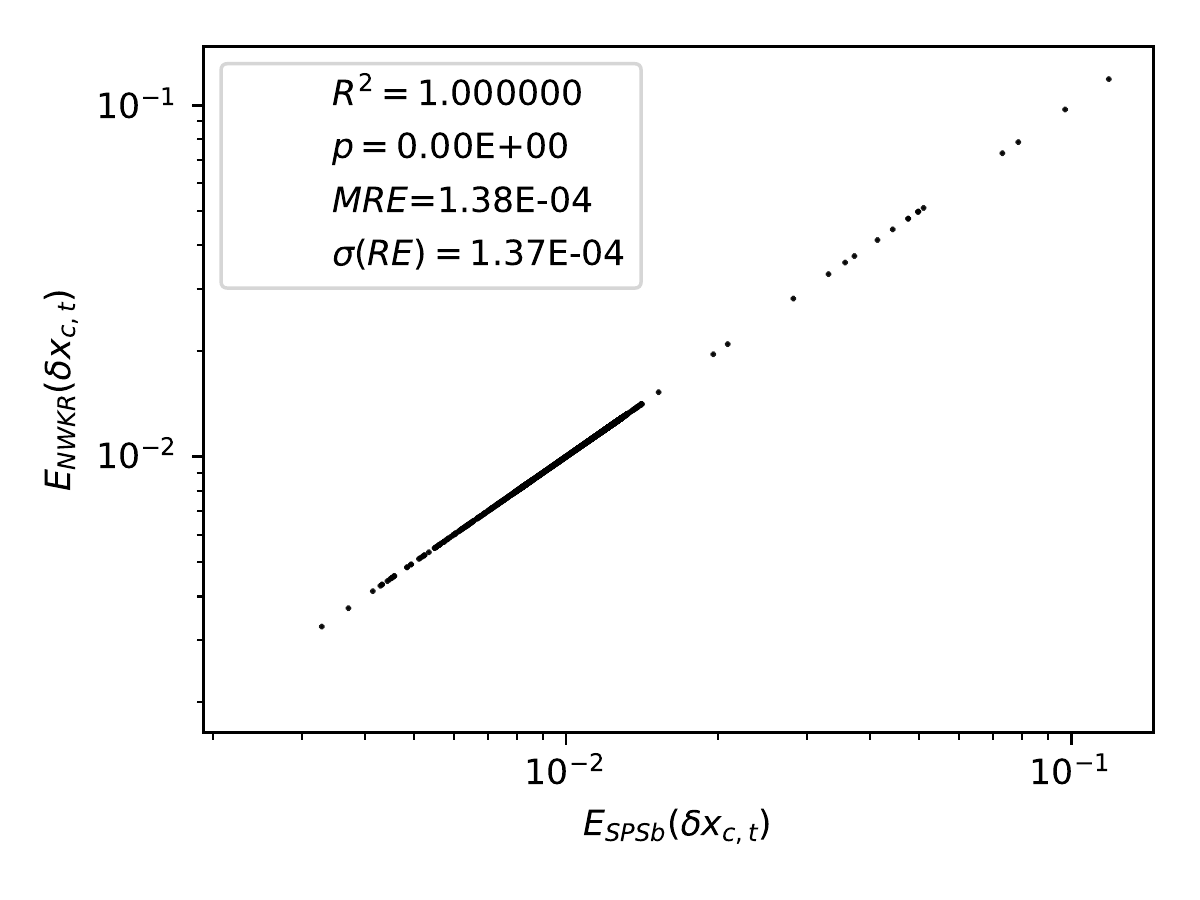}}
    \subfloat[]{
        \includegraphics[width=.49\textwidth]{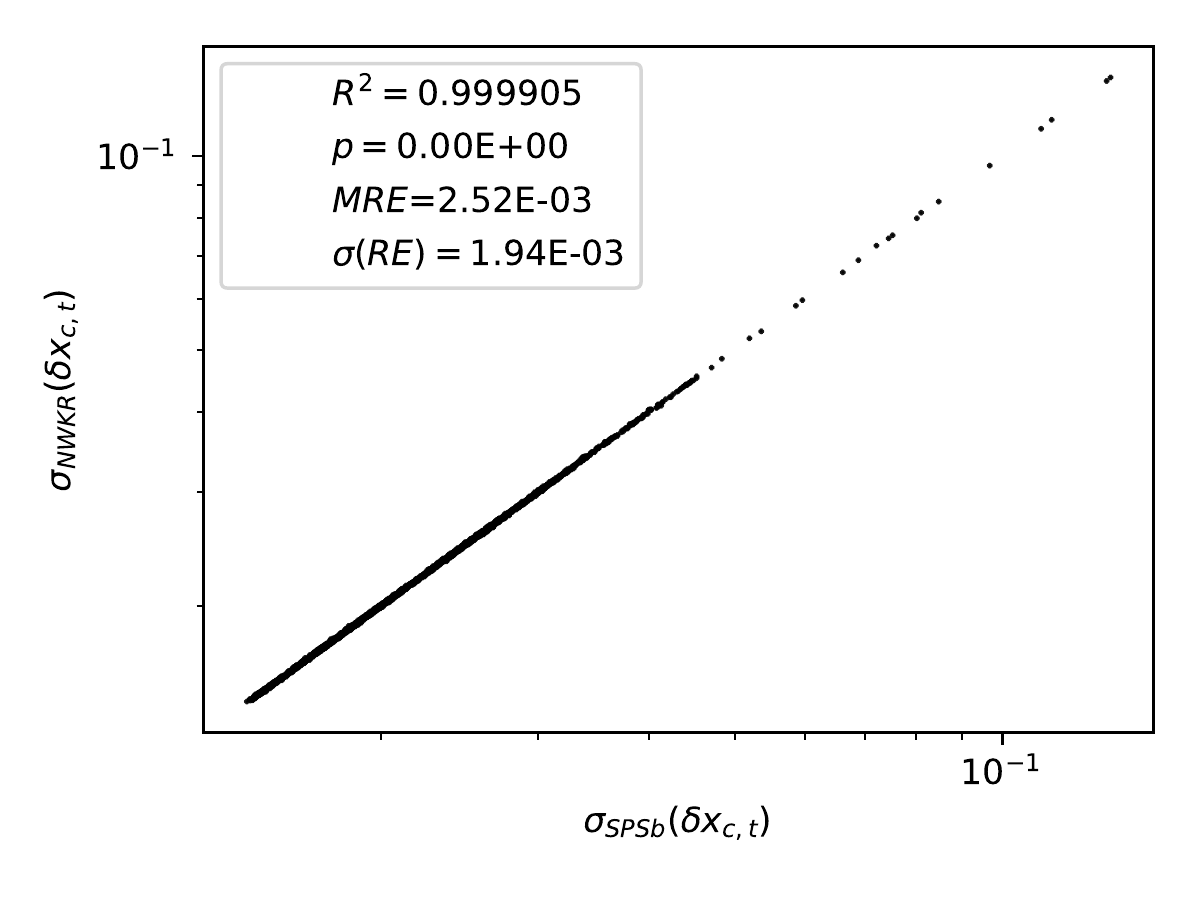}}
    \caption{\paneldeflong{a}{left} For all possible predictions to be made on the plane, a comparison of the expectation values obtained with \spsb\ at $\scinot{5}{5}$ bootstrap cycles and \nwkde. The match is, for all prediction purposes, perfect. In the legend, we report the value of $R^2$ for the observations, as well as the p-value for a linear regression (which is below machine precision, so it approximates to 0), mean relative error (the absolute value of differences normalized), and the standard deviation of the relative error. \\
    \paneldeflong{b}{right} For all possible predictions to be made on the plane, a comparison of the standard deviations obtained with \spsb\ at $\scinot{5}{5}$ bootstraps and \nwkde. The match is, again, perfect for prediction purposes.}
    \label{fig:convergence:comparison}
\end{figure}

We computed expectations and standard deviations for economic complexity data with both \spsb\ ($\scinot{5}{5}$ bootstraps) and \nwkde. The results here refer to the calculation for \gdp\ prediction, but the same results are obtained with products predictions. It can be clearly seen from \reffigure{fig:convergence:convergence_main}\refpanel{a} that the expectation values for \spsb\ converge to \nwkde\ expectation values as the number of bootstraps increases. We show that the mean average error $\text{MAE}[E_{\text{\spsb}}(\delta x)] = \text{abs} \left[ \frac{E_{\text{\spsb}}(\delta x)-E_{\text{\nwkde}}(\delta x)}{E_{\text{\nwkde}}(\delta x)} \right]$ converges numerically to zero (by $E_M(\delta x)$ we mean the expectation value of the displacement of $x$ calculated with method $M$). The standard deviations converge as well, as can be seen from \reffigure{fig:convergence:convergence_main}\refpanel{b}. Here too we calculate $\text{MAE}[\sigma_{\text{\spsb}}(\delta x)] = \text{abs} \left[ \frac{\sigma_{\text{\spsb}}(\delta x)-\sigma_{\text{\nwkde}}(\delta x)}{\sigma_{\text{\nwkde}}(\delta x)} \right]$. A comparison of the values obtained for expectations with the two methods is shown in \reffigure{fig:convergence:comparison}\refpanel{a}. The difference between predictions with the two methods is $~\scinot{3}{-5}$ on average with a standard deviation of $~\scinot{3}{-5}$. A comparison of the standard deviations obtained with the two methods is shown in \reffigure{fig:convergence:comparison}\refpanel{b}. The difference between the two methods in this case is $~\scinot{6}{-4}$ on average with a standard deviation of $~\scinot{5}{-4}$. For the purpose of GDP prediction we can therefore say that the two methods are completely interchangeable. The time complexity for \spsb\ is of the order $O(NB)$, while for \nwkde\ is $O(N)$, so with $B=1000$ bootstraps (as reccommended by the literature\cite{Tacchella2018}) \nwkde\ is expected to be $~1000$ times faster. The same is not true for space complexity, since the original \spsb\ can be implemented with $O(N)$ memory requirements like \nwkde.\\

\begin{figure}[H]
    \centering
    \subfloat[]{
        \includegraphics[width=.49\textwidth]{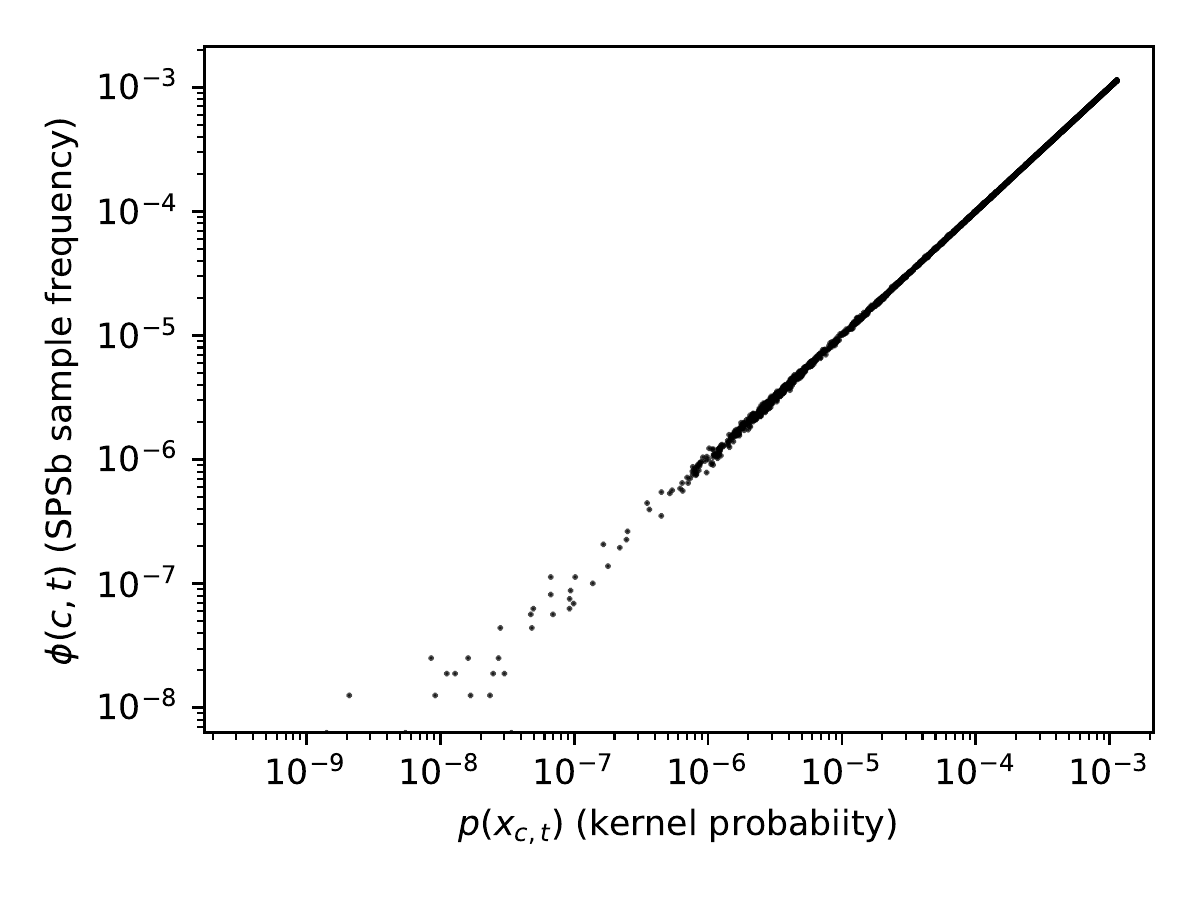}}
    \subfloat[]{
        \includegraphics[width=.49\textwidth]{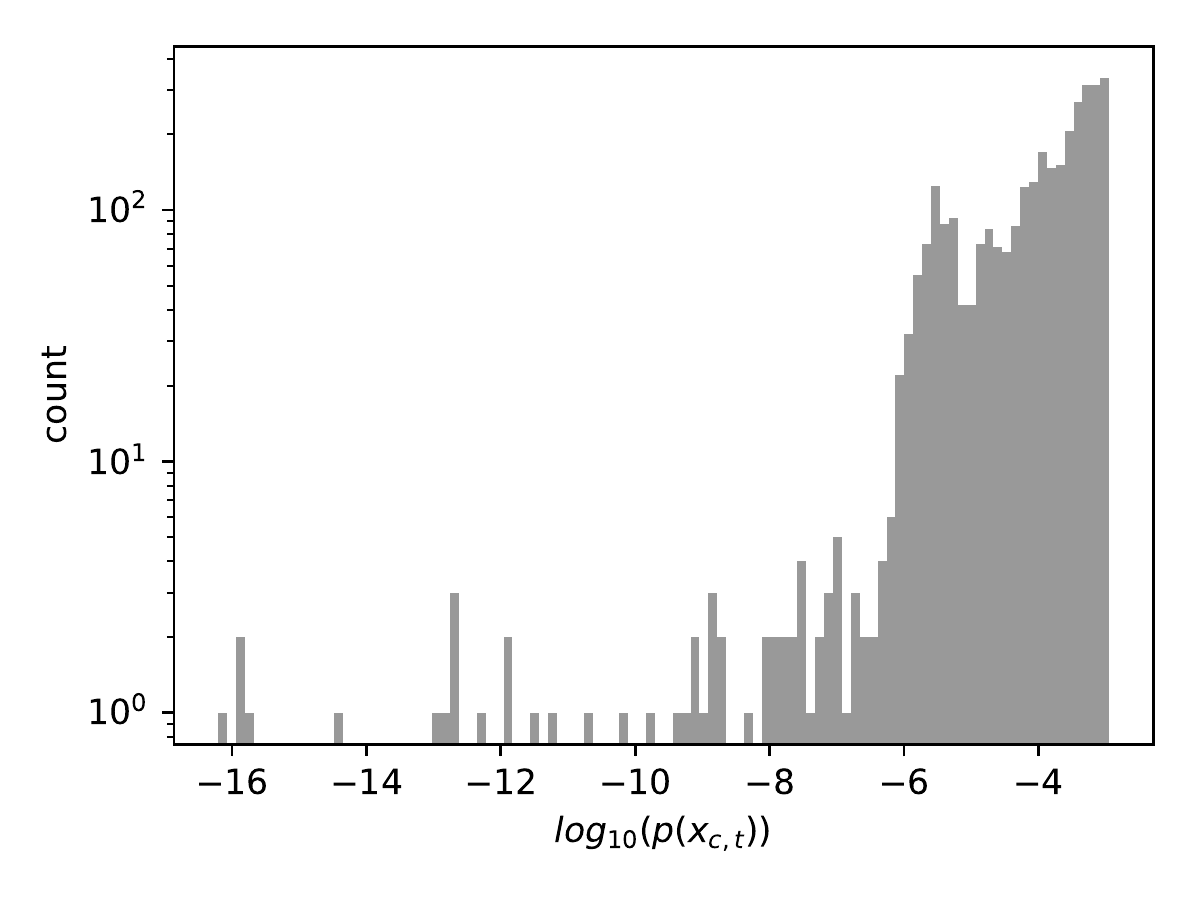}}
    \caption{\paneldeflong{a}{left} Sample frequencies $\phi(c,t)$ converge to kernel probabilities $p(c,t)$, as defined in \refeq{eq:kernel_prob_phi}. This plot compares them after $B=\scinot{5}{4}$ bootstrap cycles of \spsb\ (with $N=100$, i.e. $\scinot{5}{6}$ sampled analogues). The values, as expected, start to visibly diverge around $10^{-6}$.\\
    \paneldeflong{b}{right} Histogram of the probabilities assigned by the kernel to all analogues on the plane, for a typical prediction. It can be seen that a sizeable proportion of the analogues has probability e.g. $\leq 10^{-5}$. They will therefore not be included in \spsb\ if the number of analogues sampled is of order $~10^{5}$.}
    \label{fig:convergence:kernel_prob}
\end{figure}

The convergence does not reach machine precision even at $\scinot{5}{5}$ bootstrap cycles of \spsb\ because many of the analogues have extremely small probabilities to appear in a bootstrap. In Figure \ref{fig:convergence:kernel_prob}\refpanel{b} we show the probabilities assigned by the kernel to all analogues of the plane for a typical prediction. In \reffig{fig:convergence:kernel_prob}\refpanel{a} we compare, for a typical prediction, the sample frequency of each analogue with the sampling probability assigned to it by the kernel. It can be clearly seen from both figures that a sizeable proportion of the analogues has no chance to appear even in a bootstrap of $\scinot{5}{5}$ cycles since about 30 per cent of them have probability significantly $\leq 10^{-7}$ (each bootstrap samples $N=10^2$ analogues). These analogues are instead included in the \nwkde\ estimate, although with a very small weight. To obtain complete convergence one would have to sample, in total, as many analogues as the inverse of the smallest probability found among the analogues, and this number can go up to $10^{25}$ in typical use cases. We expect the discrepancies to decrease with the total number of samples (i.e. $NB$), as more and more analogues are sampled with the correct frequency. A visual representation of such discrepancies can be seen in \reffig{fig:convergence:kernel_prob}\refpanel{a}, where we plot the kernel probabilities of each available analogue $p(c,t)$ against the sampled frequencies $\phi(c,t)$ for a bootstrap of $\scinot{5}{4}$ samples. Discrepancies start to show, as expected, at a probability of about $10^{-6}$.

\subsection{\hmm\ regularization reduces noise and increases nestedness}
\label{subsection:regularization_effect}

In analogy to what happens for countries, product classes too can be represented as points $(\lp_t,C_t)$ on the \clplong\ (\clp) plane. Their trajectories over time $t$ can be then considered, and one can find the average velocity field $\vfield$ by dividing the \clp\ plane into a grid of square cells and averaging the time displacements $(\delta \lp_t,\delta C_t)$ of products per cell\footnote{The procedure of averaging per cell on a grid can be considered a form of nonparametric regression, but it is by no means the only technique available to treat this problem. All the following results hold independently of the regression technique used to do the spatial averages, as reported in \cite{Angelini2017}}.\newline
The product model described in \cite{Angelini2017} and summarised in \refsection{subsection:clpmotion} explains the $\vfield$ field in terms of competition maximization. For each product, it is possible to compute the Herfindahl index $H(p,t)$ (\refeq{eq:herfindahl_def} \refsection{subsection:clpmotion}), which quantifies the competition on the international market for the export of product $p$ in year $t$. The lower $H(p,t)$, the higher the competition. Averaging the values of $H(p,t)$ per cell on the \clp\ plane gives rise to a scalar field, which we call the Herfindahl field $H$. The inverse of the gradient of this field $-\nabla H$ explains the average velocity field (\refeq{eq:hfield-grad-def}, \refsection{subsection:clpmotion}), much like a potential.\newline

The original work where this model was proposed used a dataset of about 1000 products, classified according to the Harmonized System 2007 \cite{Angelini2017}. The Harmonized System classifies products hierarchically with a 6-digit code. The first 4 digits specify a certain class of product, and the subsequent two digits a subclass (see \refsection{subsection:datasets}). In \cite{Angelini2017}, the export flux was aggregated at the 4 digit level, and we will refer to this dataset as \noreg{4}. We recently obtained the full 6-digit database, comprehensive of about 4000 products. We calculated the model on $\mcp$ matrices at 6 digit level (\noreg{6}), to compare it with the \noreg{4} case. We also obtained the same 6-digit dataset regularized with the aforementioned \hmm\ method \cite{Tacchella2018} (see \refsection{subsection:hmm}), which we will call \reg{6}. This method goes beyond the classical definition of the $\mcp$ matrix as a threshold of the \rca\ matrix (\refeq{eq:mcp_definition},\ref{eq:rca_definition}, in \refsection{subsection:clpmotion}). Because the value of \rca\ fluctuates over time around the threshold, it can lead to elements of the $\mcp$ matrix switching on and off repeatedly, polluting the measurements with noise. The \hmm\ algorithm stabilizes this fluctuation. Because of this, it can significantly increase the accuracy of GDP predictions \cite{Tacchella2018}.\newline

We computed the \clp\ motion model on the three different datasets hitherto described. The results can be compared visually in \reffigure{fig:fields_gradients}. Each of the panels in \reffig[s]{fig:fields_gradients}\refpanel{a,c,e} show the $\vfield$ for one of the datasets, and the corresponding panels \refpanel{b,d,f} plot the $H$ field in colors, and the gradient $-\nabla H$ as arrows. The yellow line superimposed on each of the $\vfield$ plots is the minimum of the vertical component of the velocity field along each column of the grid on the plane, together with error bars obtained via bootstrap. The blue line superimposed on each of the $H$ plots is the minimum of the $H$ field along each column of the grid together with error bars.\newline

\subsubsub{Noise reduction} Panels in \reffig[s]{fig:fields_gradients}\refpanel{a-b} are almost identical to those in \cite{Angelini2017}, since the \noreg{4} data set is the same with the addition of one more year of observations (namely 2015). \reffig[s]{fig:fields_gradients}\refpanel{c-d} represent the velocity and Herfindahl field obtained with \noreg{6}. The most noticeable change is the strong horizontal component of the velocity field: Complexity changes much faster than in \noreg{4}. We believe this is due to two effects. The first one is the increased noise: when a 4-digit code is disaggregated into many 6-digit codes, there are fewer recorded export trades for each product category. This means that each individual 6-digit product category will be more sensitive to random fluctuations in time, of the kind described in \refsection{subsection:hmm}. The second source of change is due to overly specific product classes. There are some products, such as e.g. products typical of a specific country, for which we would expect generally low Complexity. It typically happens that these products are exported by almost only one, fairly high-Fitness, country, which produces it as a speciality. When the Complexity of such products is computed with \refeq{eq:fitcompmap} (\refsection{subsection:fitcomp_def}), it will be assigned a high value, because they have few high-quality exporters. This effect increases the Complexity of the product and is stronger in more granular data. Combined with the stronger fluctuations coming from disaggregation, it contributes to noise in the Complexity measurements.\newline

Another, stronger argument in favour of noise causing fast Complexity change over the years in \noreg{6} is \reffig[s]{fig:fields_gradients}\refpanel{e-f}. These figures show the velocity and Herfindahl field for the regularized \reg{6} data. It is clear that the horizontal components of the $\vfield$ field are much smaller compared to \noreg{6}, and that the only change in the data comes from the regularization, which was explicitly developed to reduce the impact of random fluctuations in export measurements. We, therefore, conclude that the \hmm\ regularization is effective in reducing noise and generating smoother Complexity time series. Another interesting observation is that the $\vfield$ obtained from \reg{6} is very similar to the \noreg{4} one. Therefore we would like to conjecture that aggregating data from 6 digits to 4 has an effect similar to that of reducing noise with the \hmm\ algorithm. We will see in the next section that there is a further evidence to this conjecture.\newline

\subsubsub{Increase in nestedness} A yet undocumented effect of \hmm\ regularization is the increase in nestedness of the $\mcp$ matrices. It can be visualized by looking at \reffigure[s]{fig:nestedness_kde}\refpanel{a,c,e}. Here we show a point for each nonzero element of all $\mcp$ matrices available in each dataset. To be able to resolve the differences in density, we computed a kernel estimate of the density of points on the plane. The horizontal axis is the value of rank(Complexity), while rank(Fitness) is on the vertical axis. All three datasets feature very nested matrices, as expected, but \reg{6} has one peculiarity. The top left corner of \reffig{fig:nestedness_kde}\refpanel{e} exhibits in fact a higher density than the other two. This means that regularization has the effect of activating many low-Complexity exports of high-Fitness countries. This makes sense since we expect the thresholding procedure described in \refsection{subsection:fitcomp_def} to be noisier in this area. Indeed, we know that the high-Complexity products are exported only by high-Fitness countries, so we expect the numerator of the \rca$\cp$ (proportional to the importance of $p$ in total world export, see \refsection{subsection:fitcomp_def}) in this area to be small. We also know \rca$\cp$ is proportional to the importance of product $p$ relative to total exports of $c$, so we expect it to be high in the low-Complexity/low-Fitness area since low-Fitness countries export few products. Furthermore, it has been described in \cite{Angelini2017} that countries are observed to have similar competitive advantage in low-Complexity products regardless of their level of Fitness. So in the high-Fitness/low-Complexity area, we expect to observe a lower numerator, possibly fluctuating around the thresholding value, due to the high diversification of high-Fitness countries.\newline

A higher density in the high-Fitness, low-Complexity area naturally results in more nested matrices. To show this, we computed the well-known NODF \cite{Almeida-Neto2008,Beckett2014,pyfalcon} measure of nestedness for all $\mcp$ matrices in all datasets. The results can be found in \reffigure{fig:nestedness_measures}\refpanel{a}, and show clearly that \reg{6} matrices are much more nested than unregularized ones. Another observed result is that \noreg{4} matrices are slightly but consistently more nested than the \noreg{6} ones. This is further support for our conjecture that aggregating from 6 to 4 digit has an effect similar to regularizing with an \hmm\ model. \reffigure{fig:nestedness_measures}\refpanel{b} shows the significance level of the NODF measurements. In order to assess significance, we computed $n_{\text{obs}}$, the observed value of NODF on the $\mcp$ matrices, and we compared it with $n_{\text{null}}$ the NODF obtained from null models. The null models usually generate new adjacency matrices at random while holding some of the properties of the observed matrix (such as e.g. total number of nonzero elements) fixed. This is a way to control for the effect of the fixed property on the nestedness. Several runs of a null model generate an empirical probability distribution $p(n_{\text{null}})$. The p-value of the measurement is assessed by calculating in which quantile of $p(n_{\text{null}})$ the observed value $n_{\text{obs}}$ falls. In \reffig{fig:nestedness_measures}\refpanel{b} we report the ratio between $n_{\text{obs}}/E_p(n_{\text{null}})$ and the scaled standard deviation of the null distribution $\sigma(n_{\text{null}})/E_p(n_{\text{null}})$, for three common null models \cite{Beckett2014}. The scaling allows to compare very different distributions on the same axis. The ratio of $\sigma(n_{\text{null}})$ to $n_{\text{obs}}-E_p(n_{\text{null}})$ is very small. Thus, the observed measurements' significance is so high that there is no need to calculate quantiles.\\

\subsubsub{Model breakdown at 6 digits} Another observation that can easily be made from \reffigure{fig:fields_gradients} is that, while it works well for 4-digit data, the model of product motion has trouble with reproducing the data at the 6-digit level. Regressing the $\vfield$ components against the derivatives of the $H$ field, as shown in \reftab{table:r2_fields}, seems to indicate that the 6-digit models work better\footnote{However, the 4-digit BACI dataset \reg{4} has one peculiarity that needs explaining. Specifically, the bottom right corner of \reffig{fig:fields_gradients}\refpanel{b} does not contain the maximum of $H$ that is found in all other datasets ever observed (including the Feenstra dataset studied in \cite{Angelini2017}). This causes the gradient of $H$ in that area to produce small values, which do not match the high vertical components of $\vfield$ in the same spot, significantly lowering the $R^2$ coefficient of a linear regression.}. But one key feature of the model disappears when moving from 4 to 6 digits. The yellow and blue lines in \reffig{fig:fields_gradients} indicate a kernel regression of respectively the minima of the $\vfield$ field and the minima of the $H$ field across each column of the grid (together with error bars obtained via bootstrap). The model predicts that $\vfield$ will be almost zero where the minima of $H$ lie, but at 6 digits this feature disappears, and the minima lines become incompatible with each other. We are currently lacking an explanation of this behaviour, that seems independent of regularisation.\\

\begin{table}
\caption{$R^2$ coefficients of a linear regression of $\vfield$ components against the derivatives of the $H$ field along the x-axis (Complexity) and y-axis (\logprody).}

\small 
\centering
\begin{tabular}{ccc}
\textbf{dataset}    & \textbf{y-axis}    & \textbf{x-axis}\\
4-digit non-regularized        &0.103        &0.023\\
6-digit non-regularized        &0.487        &0.200\\
6-digit regularized            &0.558        &0.135\\
\end{tabular}
\label{table:r2_fields}
\end{table}

\subsection{Predictions on products with \spsb}
\label{subsection:predictions}
Dynamics of products on the \clp\ plane appears to be laminar everywhere, in the sense that the average velocity field seems to be smooth\cite{Angelini2017}, similarly to what happens to countries on the Fitness-\gdp\ plane \cite{Cristelli2013}. If so, then it's a reasonable hypothesis that the information contained in the average velocity field can be used to predict the future positions of products on the plane.
We tried to predict the future displacement of products with \spsb. Because the number of products is about 1 order of magnitude larger than the number of countries used in \cite{Tacchella2018}, the computational demand of the algorithm induced us to develop the proof of convergence reported in \refsection{subsection:convergence}.\newline

The results for the backtests on this methodology are reported in \reffigure{fig:predictions:predictions}. We predicted the Percentage Compound Annual Growth Rate (CAGR\%) for each of the two metrics, and defined the error as $E = |\text{CAGR\%}_{\text{observed}} - \text{CAGR\%}_{\text{forecasted}}|$, so that if e.g. Complexity increases by 2\% and we forecast 3\%, $E=1\%$. The forecasts are made at timescales $\Delta t = 3,4,5$ years. We used the three datasets \reg{6}, \noreg{6} and \noreg{4}. The predictions are not very accurate, with an error between 12\% and 6\% for \logprody\ and in the 32-13\% range for Complexity. We compared the predictions to a \define{random baseline}, i.e. predicting the displacement by selecting an observed displacement at random from all the available analogues. Compared to the random baseline, \spsb\ is always more accurate. One peculiarity about the predictions, though, is that they are generally much smaller in magnitude than the actual displacements observed. This led us to add another comparison, which we call \define{static baseline}, that consists in predicting zero displacement for all products. Compared to this baseline, \spsb\ still systematically shows some predictive power for \logprody, especially in \noreg{4}, but is definitely worse when predicting Complexity. We will clarify our explanation for this behaviour with an analogy. While the average velocity field $\vfield$ exhibits laminar characteristics, in the sense that it is relatively smooth, the actual motion of the underlying products is much more disorderly. In a given neighbourhood of the \clp\ plane, products generally move in every direction, often with large velocities, even though the average of their displacements is nonzero and small. We could tentatively describe this as a Brownian motion with a laminar drift given by $\vfield$. So trying to predict the future position of a product from their aggregate motion would be similar to trying to predict the position of a molecule in a gas. That's why the static prediction is better than a random prediction: in general, the last position of a product is a better predictor than a new random position on the plane, since the new one might be farther away. To test this Brownian motion with drift hypothesis, we added a third baseline, which we call \define{autocorrelation baseline}. It consists in forecasting the displacement of a product to be exactly equal to its previous observed displacement. If the hypothesis is true, we expect each product displacement to be uncorrelated with its displacement at previous time steps. For \logprody\ the autocorrelation baseline is always worse than the static, which we interpret as a signal that \logprody\ changes are not autocorrelated. The reverse is true for Complexity: in fact, for \noreg{4} and \reg{6} the autocorrelation baseline is the best predictor for Complexity change.\newline

As already mentioned, \spsb\ does still have slightly but systematically more predictive power than the autocorrelation and static prediction, but only for \logprody. We speculate that this is due to the fact that change in \logprody\ is actually a signal of the underlying market structure changing, as explained in \cite{Angelini2017} and in \ref{subsection:clpmotion}. The fact that this advantage over the baseline is much bigger on \noreg{4} confirms that the \logprody\ model performs significantly better on \noreg{4}, as discussed in \ref{subsection:clpmotion}. On the other hand, the autocorrelation prediction (as well as the static one) can be significantly better than \spsb\ when predicting changes in Complexity. It is not clear whether this implies that changes in Complexity are autocorrelated in time - this effect for example disappears in \noreg{6}, and will require an analysis with different techniques. But the fact that \spsb\ is always worse than the baseline, combined with the fact that regularization, which is supposed to mitigate noise, significantly reduces changes in Complexity over time raises a doubt over whether changes in Complexity are significant at all, or are drowned by noise in the $\mcp$. The fact that Complexity predictions are significantly better on the \reg{6} dataset suggests confirms the contribution of noise to Complexity changes, although it is not possible to argue that regularization is strengthening the signal coming from these changes over time, since we could not characterize any signal. This might be an important finding because it could shed some light on the nature of the Complexity metric. We suggest that an alternative line of thinking should be explored, in which one treats the Complexity of a product as fixed over time. This resonates with the data structure: product classes are fixed over the timescales considered in our analyses, and new products that might be introduced in the global market during this time are not included. It also might be derived from an interpretation of the theory: Complexity is meant to be a measurement of the number of capabilities required to successfully export a product \cite{Cristelli2013a}. Practically, this means that there is no specific reason to believe that the Complexity of (i.e. the capabilities required for) wheat, or aeroplanes, changes over the course of the 20 years typically considered in this kind of analysis. It is possible that changes in Complexity, defined as a proxy for the number of capabilities required to be competitive in a given product, occur over longer timescales, or maybe that Complexity never changes at all. If this were true, then all observed Complexity changes would be due to noise, and it would be better to consider defining a measure of Complexity that is fixed or slowly changing in time for the model. We remark that these definition problems will probably be insurmountable as long as it is impossible to give an operational definition of capabilities, and they can only be measured indirectly through aggregate proxies, i.e. countries and products. There always is a tradeoff of interpretability to pay in order to give up normative practices in favour of operational definitions, but it affects economics and social sciences more than the physical sciences.\newline

\subsection{Figures, Tables and Schemes}

\newlength{\scalereg}
\setlength{\scalereg}{.8\textwidth}
\begin{figure}[H]
    \centering
    \subfloat[\noreg{4} $\vfield$]{
        \includegraphics[clip, trim=1 1 1 41, width=.433\scalereg]{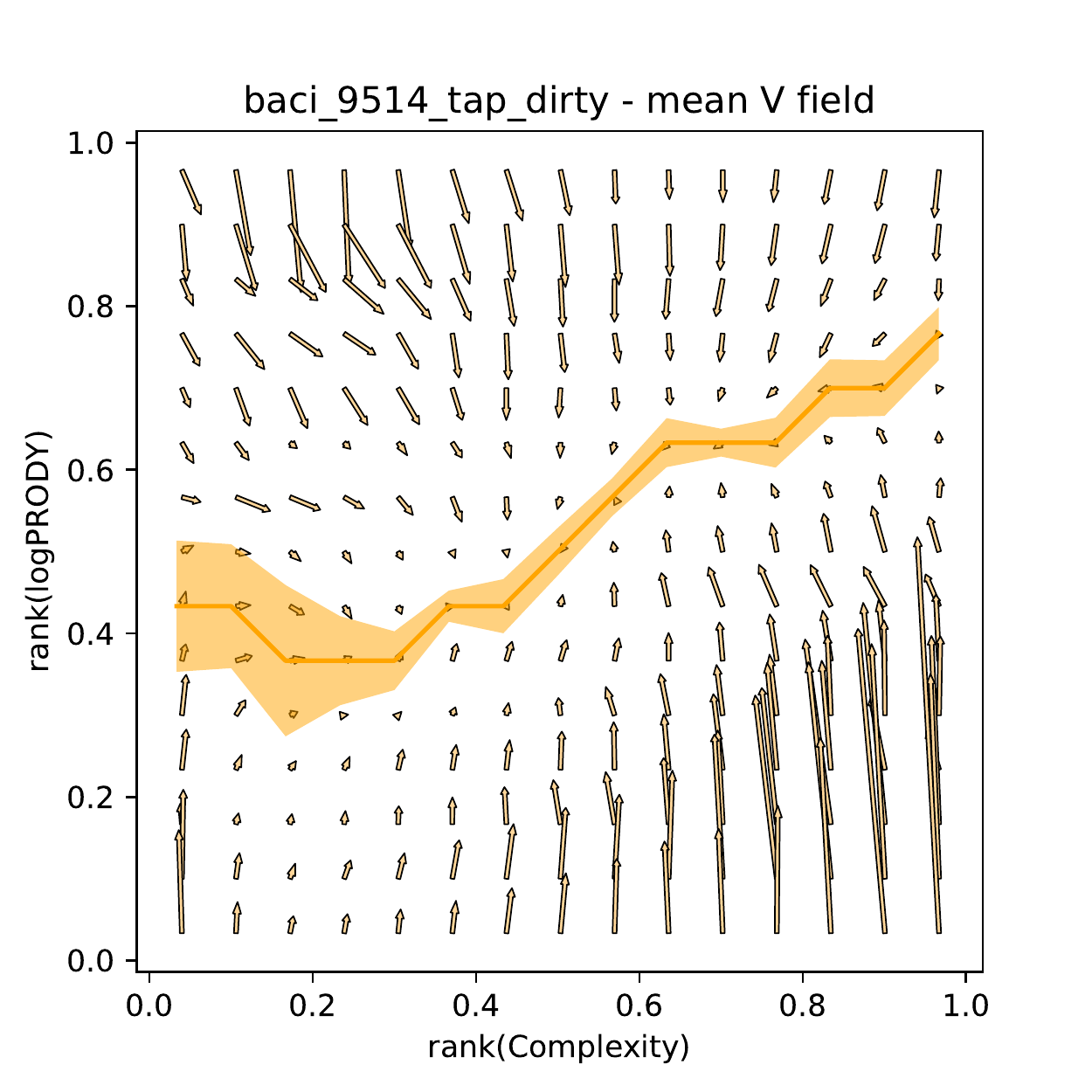}}
    \subfloat[\noreg{4} $\text{grad}(H)$]{
        \includegraphics[clip, trim=1 1 1 36, width=.567\scalereg]{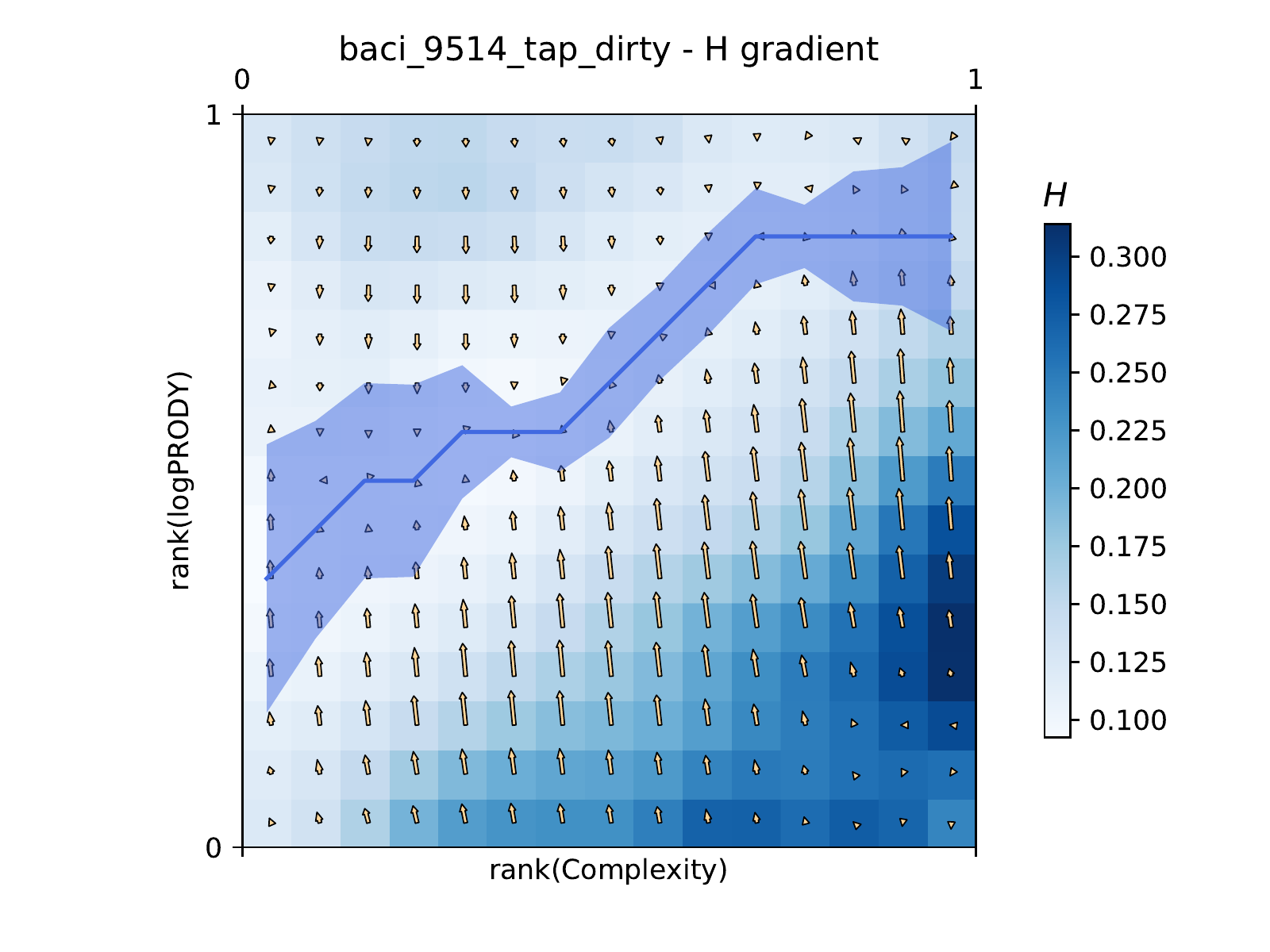}}
    \hspace{0mm}
    
    \subfloat[\noreg{6} $\vfield$]{
        \includegraphics[clip, trim=1 1 1 41, width=.433\scalereg]{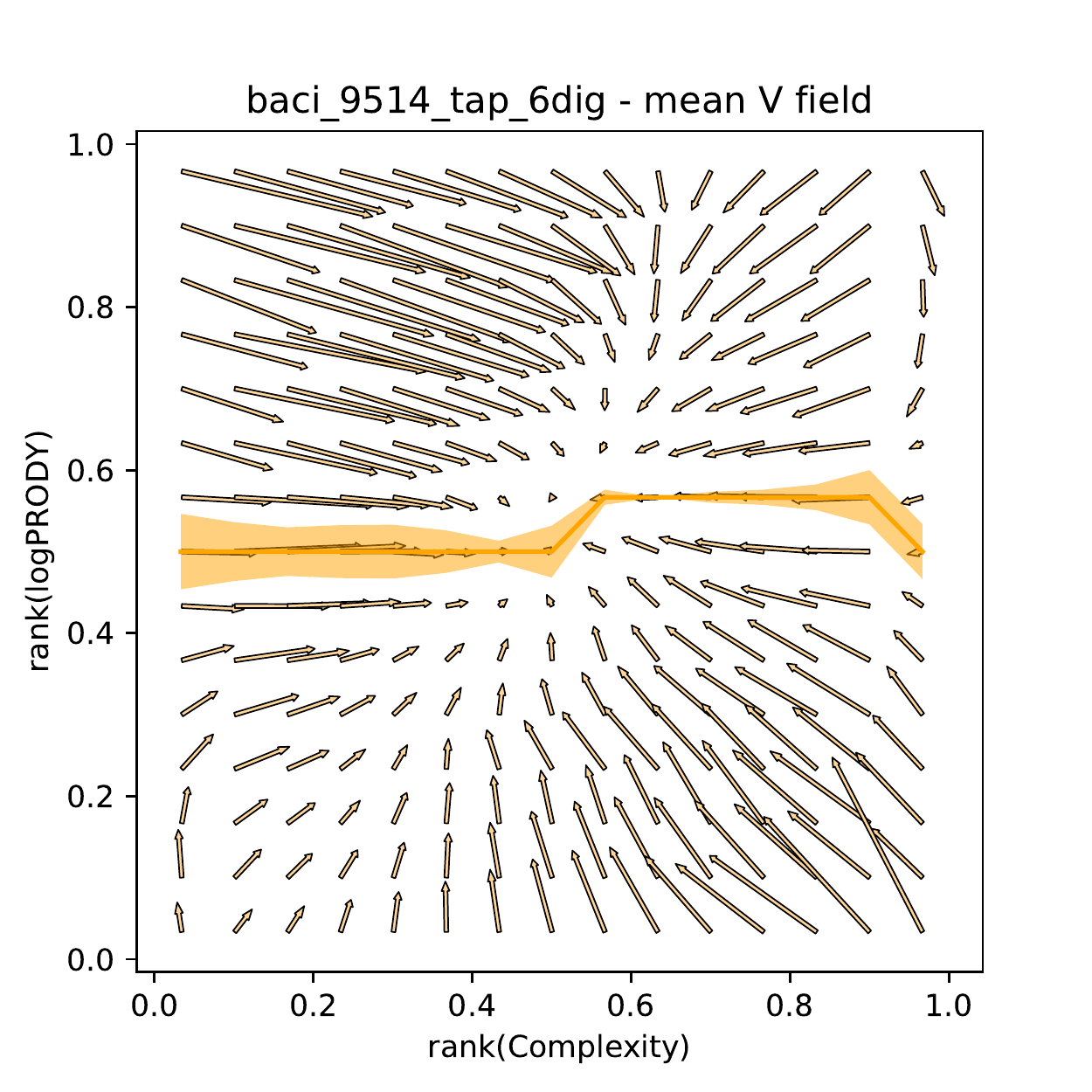}}
    \subfloat[\noreg{6} $\text{grad}(H)$]{
        \includegraphics[clip, trim=1 1 1 36, width=.567\scalereg]{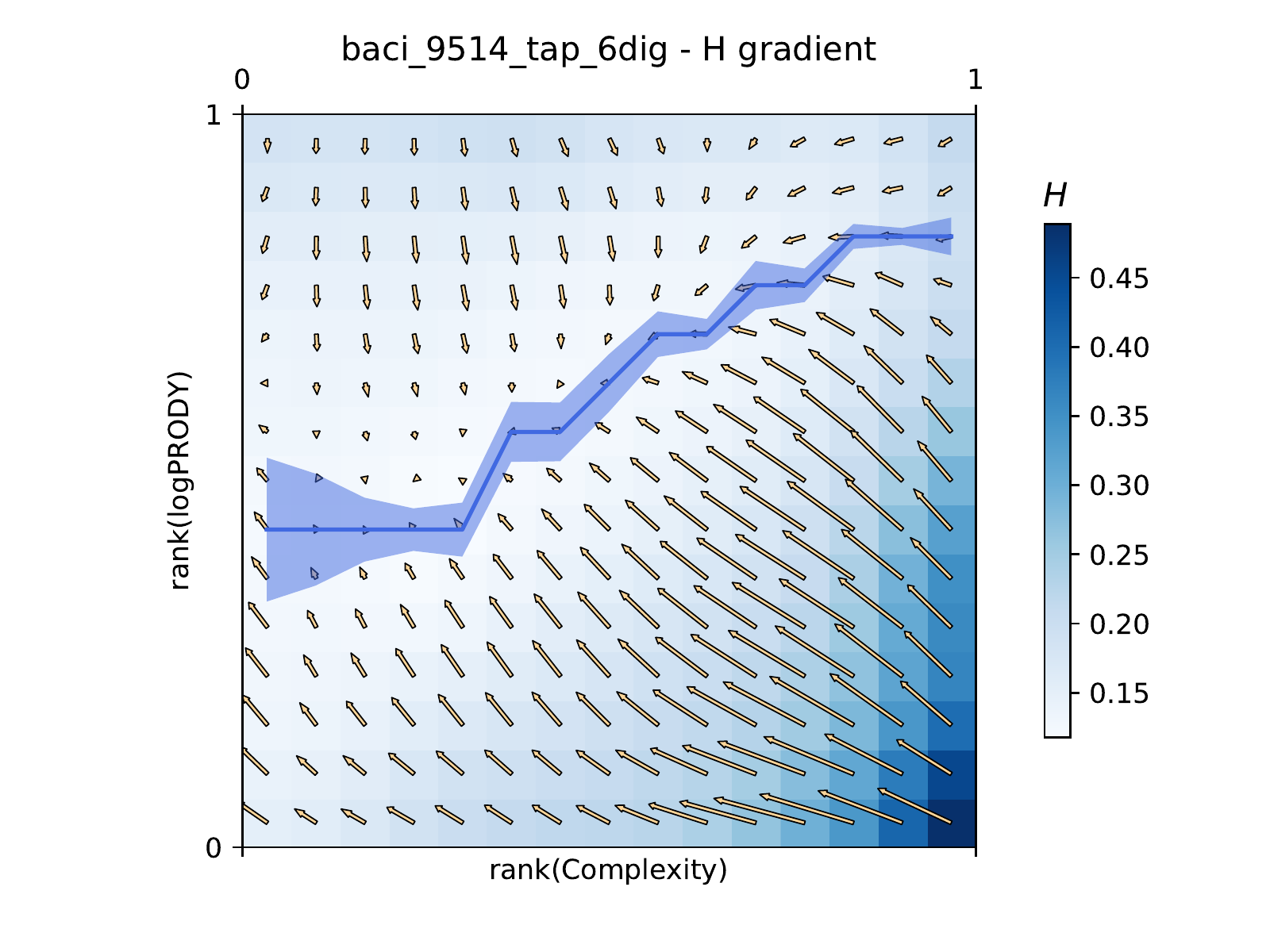}}
    \hspace{0mm}
    
    \subfloat[\reg{6} $\vfield$]{
        \includegraphics[clip, trim=1 1 1 41, width=.433\scalereg]{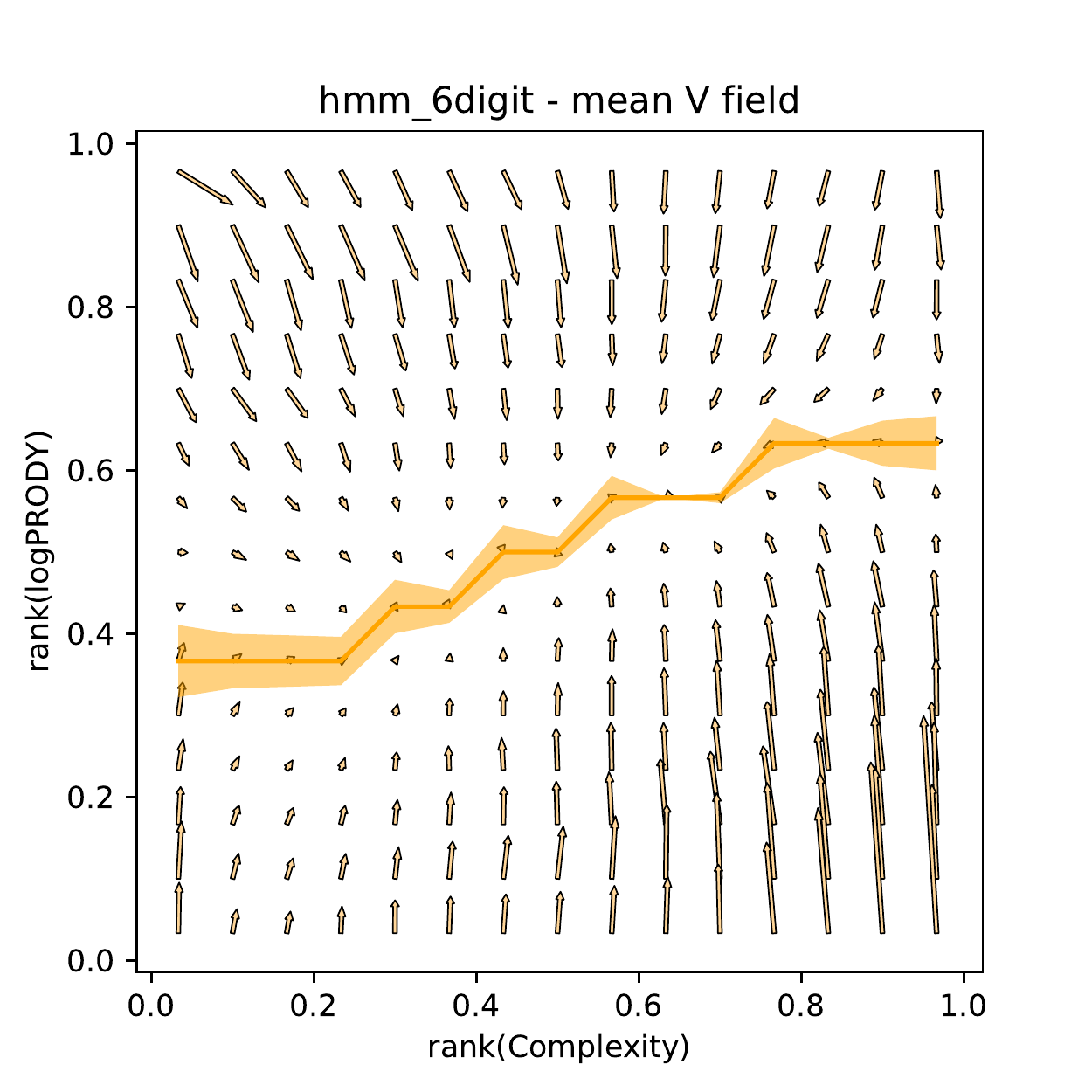}}
    \subfloat[\reg{6} $\text{grad}(H)$]{
        \includegraphics[clip, trim=1 1 1 36, width=.567\scalereg]{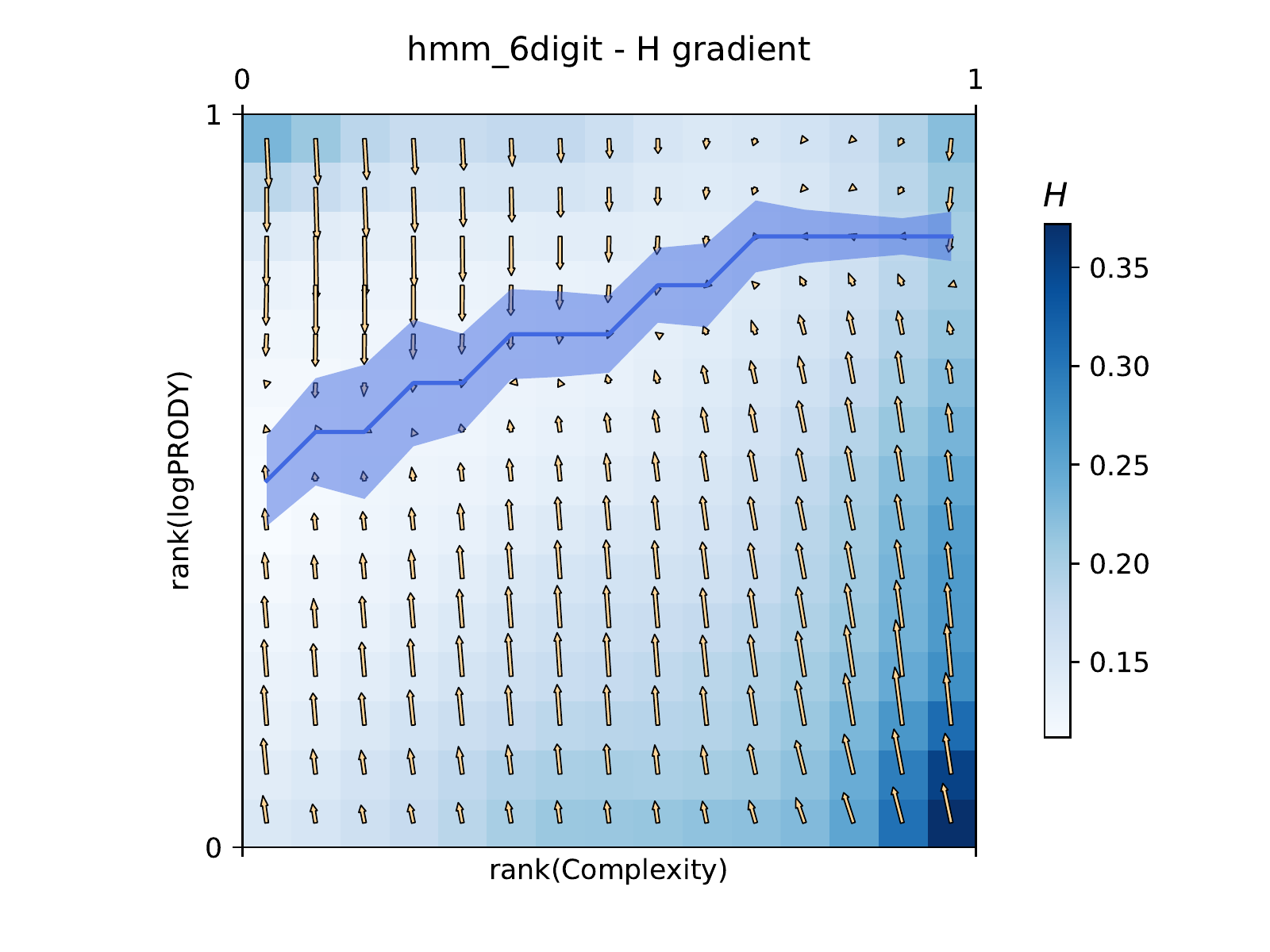}}
    \hspace{0mm}
    \caption{Comparison of the \clp\ model of motion on the different datasets used in this work. The horizontal axes mark the Complexity, and the vertical ones \logprody. Note that in these figures we use tied ranking as coordinates, instead of the observed values directly. Panels \refpanel{a,c,e} show the $\vfield$ field, together with a kernel regression of the minima of the field across the vertical direction in yellow. An uncertainty measure of this minima line has been calculated by means of a bootstrap. Panels \refpanel{b,d,f} show a heat map of the $H$ field, and its gradient. The blue line indicates the minima of the $H$ field along the vertical direction, together with an uncertainty calculated via bootstrap.\newline
    The first feature of this Figure is the difference in the $\vfield$ fields. The one calculated from \noreg{6} has much higher velocities on the Complexity axis, while the \reg{6} velocities along the same direction are much smaller. This might be an indication that much of the change in Complexity over time is actually due to noise. The second feature is that, when going from 4 to 6 digit granularity, the observed minima lines become incompatible with those predicted by the model.
    }
    \label{fig:fields_gradients}
\end{figure}

\newlength{\scalekde}
\setlength{\scalekde}{\scalereg}
\begin{figure}[H]
    \centering
    \subfloat[\noreg{4} KDE]{
        \includegraphics[width=.6\scalekde]{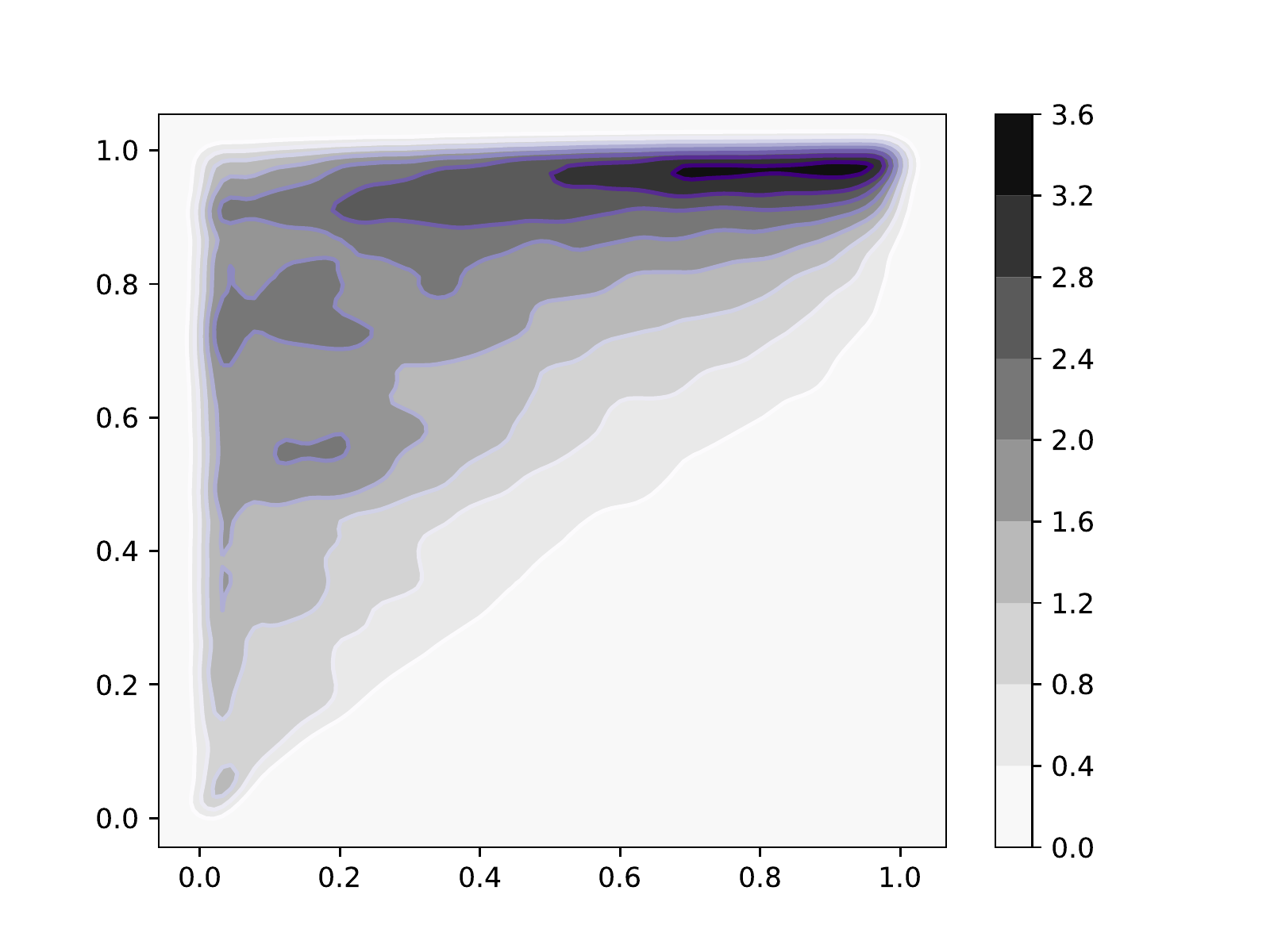}}
    \subfloat[\noreg{6} KDE]{
        \includegraphics[width=.6\scalekde]{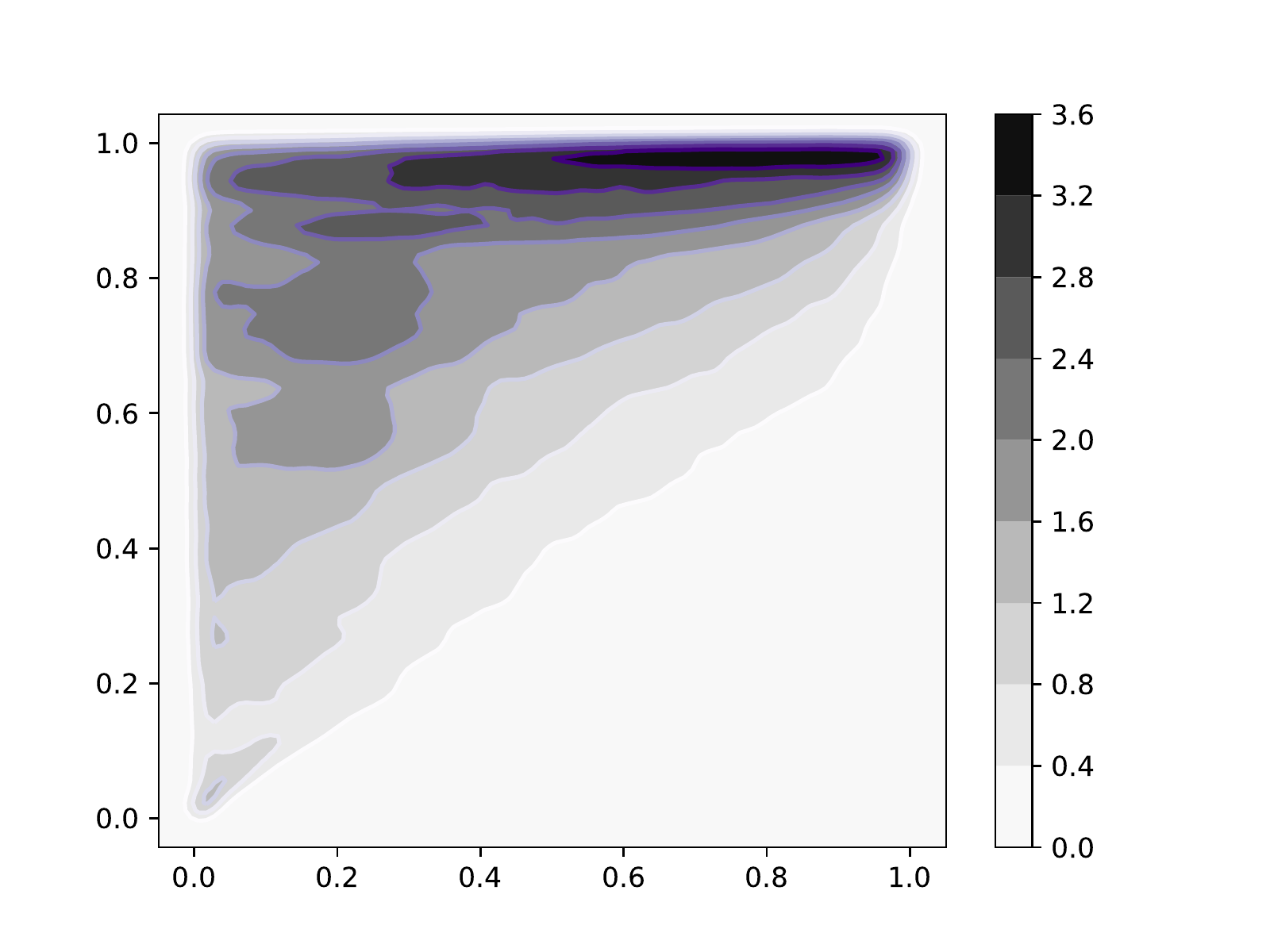}}
    \hspace{0mm}
    \subfloat[\reg{6} KDE]{
        \includegraphics[width=.6\scalekde]{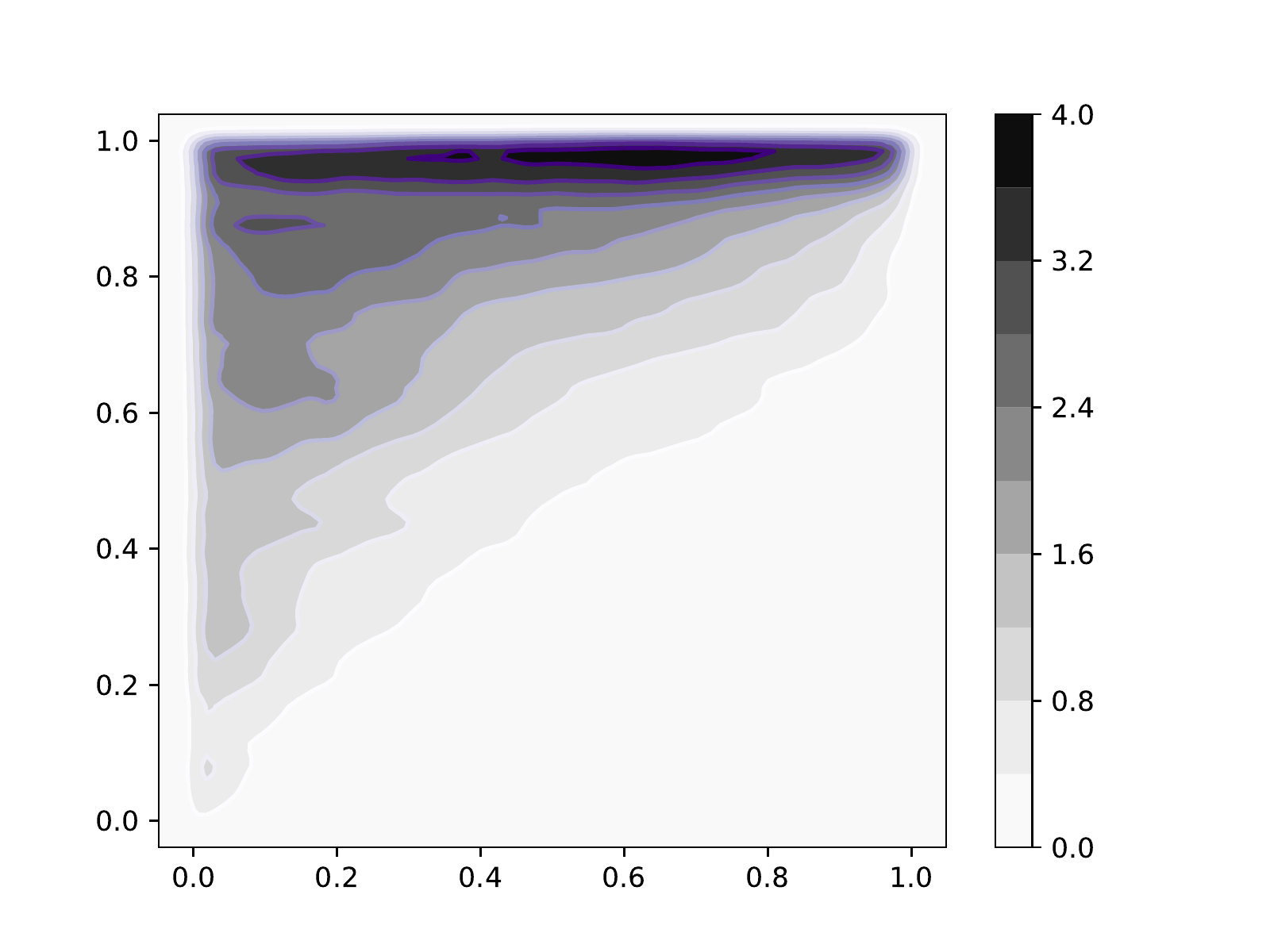}}
    \hspace{0mm}

    \caption{Comparison of $\mcp$ matrices density for the 3 datasets used in this work. In each panel, we plotted one point for each nonzero element of each $\mcp$ matrix in a dataset. Countries, ranked by increasing Fitness, are on the vertical axis, while products ranked by increasing Complexity on the horizontal axis. To be able to resolve the difference in density of points, we applied a kernel density estimate (KDE). The triangular shape suggesting nestedness is clearly visible in all three cases. The differences lie in the top left corner, where low-Complexity products exported by high-Fitness countries are found. The unregularized data  (\noreg{4}, \noreg{6}) notably have lower density here when compared with regularized matrices (\reg{6}). This is reflected in the increased nestedness of regularized matrices, as shown in \reffig{fig:nestedness_measures}.}
    \label{fig:nestedness_kde}
\end{figure}

\begin{figure}[H]
\centering
    \centering
    \subfloat[NODF measurements]{
        \includegraphics[width=.6\scalereg]{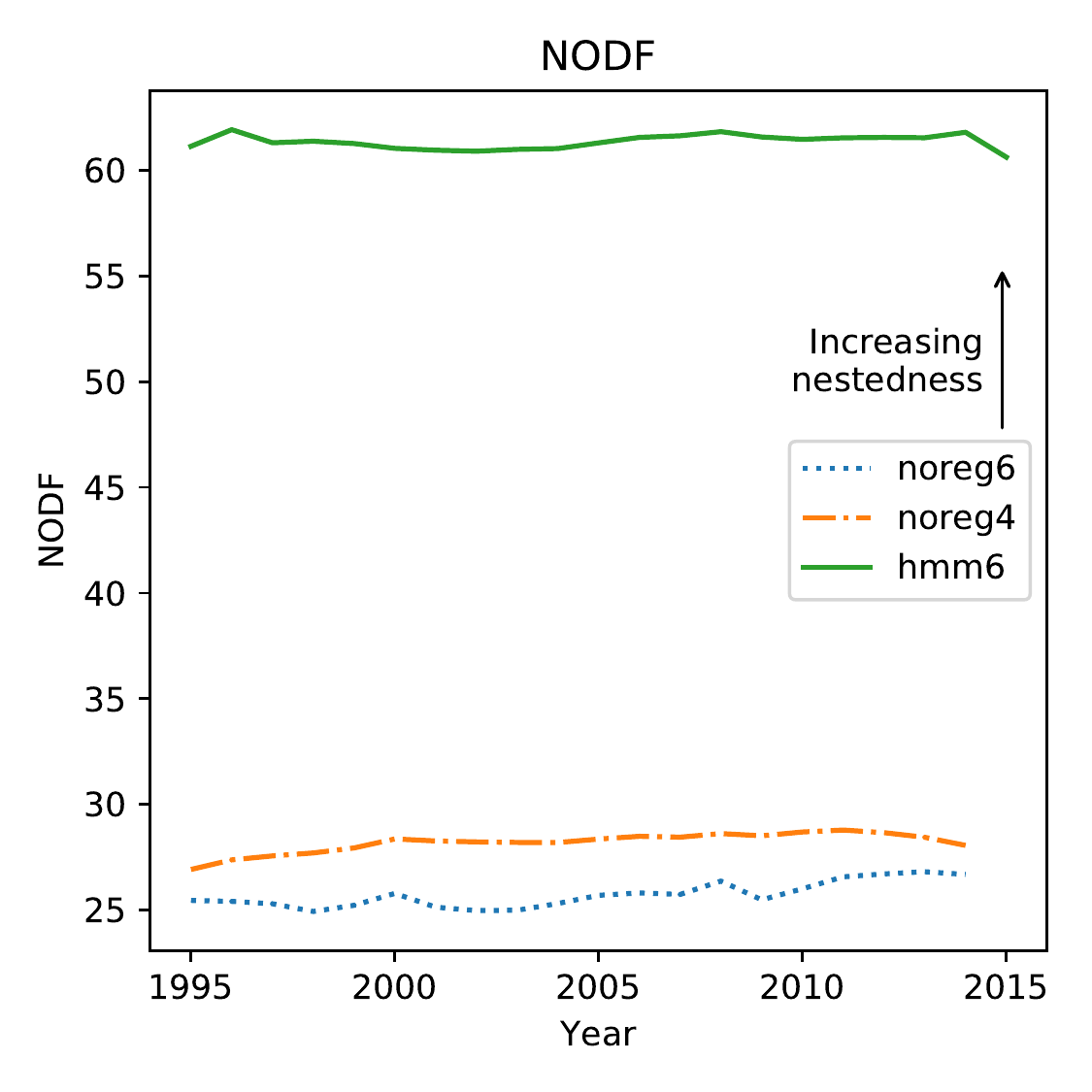}}
    \subfloat[NODF $p$-value]{
        \includegraphics[width=.6\scalereg]{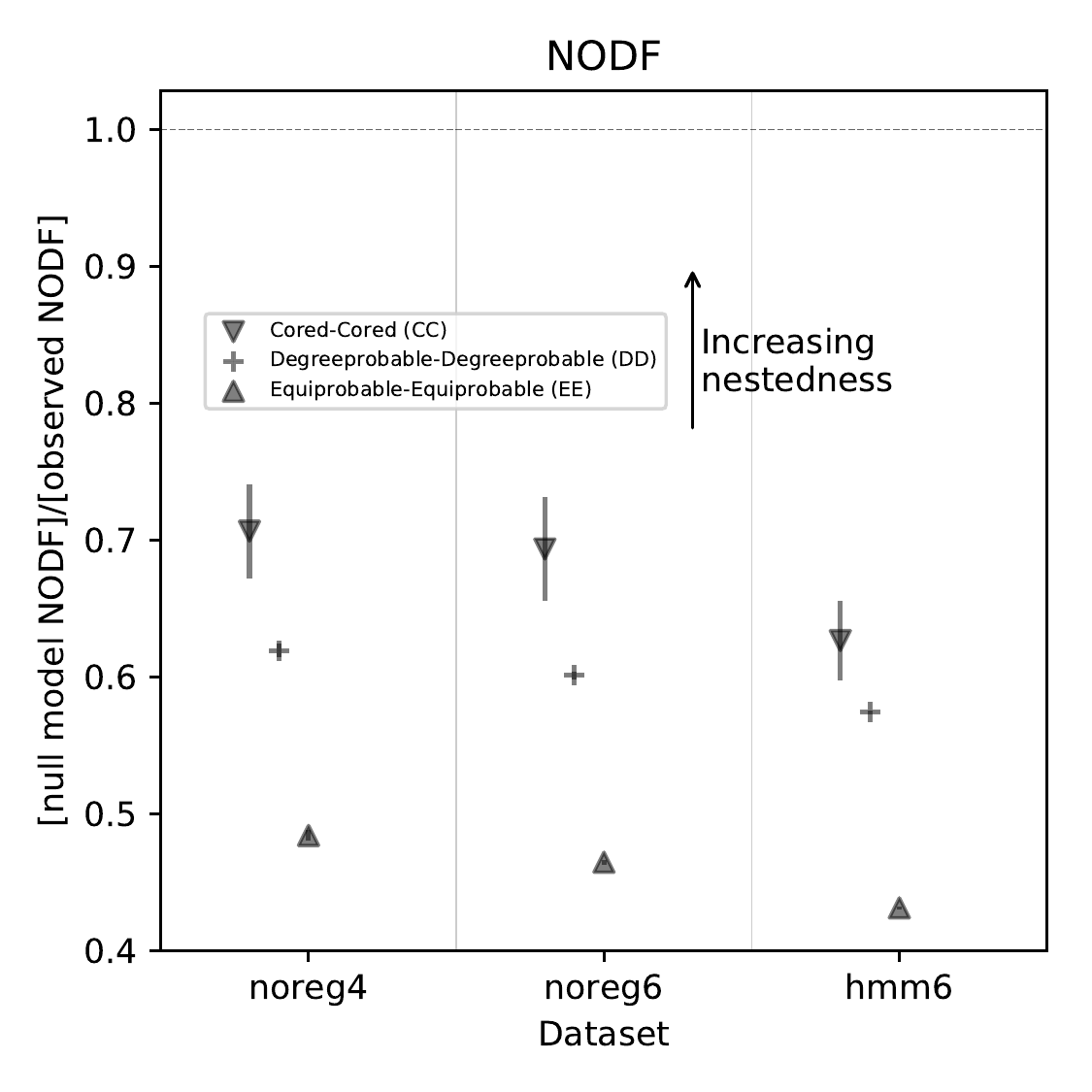}}
    \hspace{0mm}
\caption{\paneldeflong{a}{left} Measures of nestedness for the $\mcp$ matrices in the three datasets discussed in this work. We used the NODF \cite{Almeida-Neto2008} measure, which goes from 0 (no nestedness) to 100 (perfectly nested matrix). It can be clearly seen that the regularized data, \reg{6}, is much more nested than the rest, as already suggested by the observation of \reffig{fig:nestedness_kde}. The \noreg{4} dataset, though, is significantly and consistently more nested than the \noreg{6}. This suggests that aggregating from 6 to 4 digits might have a regularizing effect.\newline
\paneldeflong{b}{right} Significativity of NODF measures. We calculate an ensemble of 100 null models for each dataset and report the ratio (null model NODF)/(observed NODF). We do this for 3 commonly used null models \cite{Beckett2014}, and we report the standard deviation of the ensemble (similarly scaled) in the form of an error bar. The standard deviation of the DD and EE null models ensembles is so small that it cannot be seen in the plot. We observe that all null models have significantly smaller NODF than the observed matrices, and the results are therefore highly significant. All calculations were done with the FALCON software package \cite{Beckett2014}.
}
\label{fig:nestedness_measures}
\end{figure}

\setlength{\scalereg}{.45\textwidth}
\begin{figure}[H]
    \centering
    \subfloat[\noreg{4} Complexity predictions]{
        \includegraphics[clip, trim=11 16 14 14, width=\scalereg]{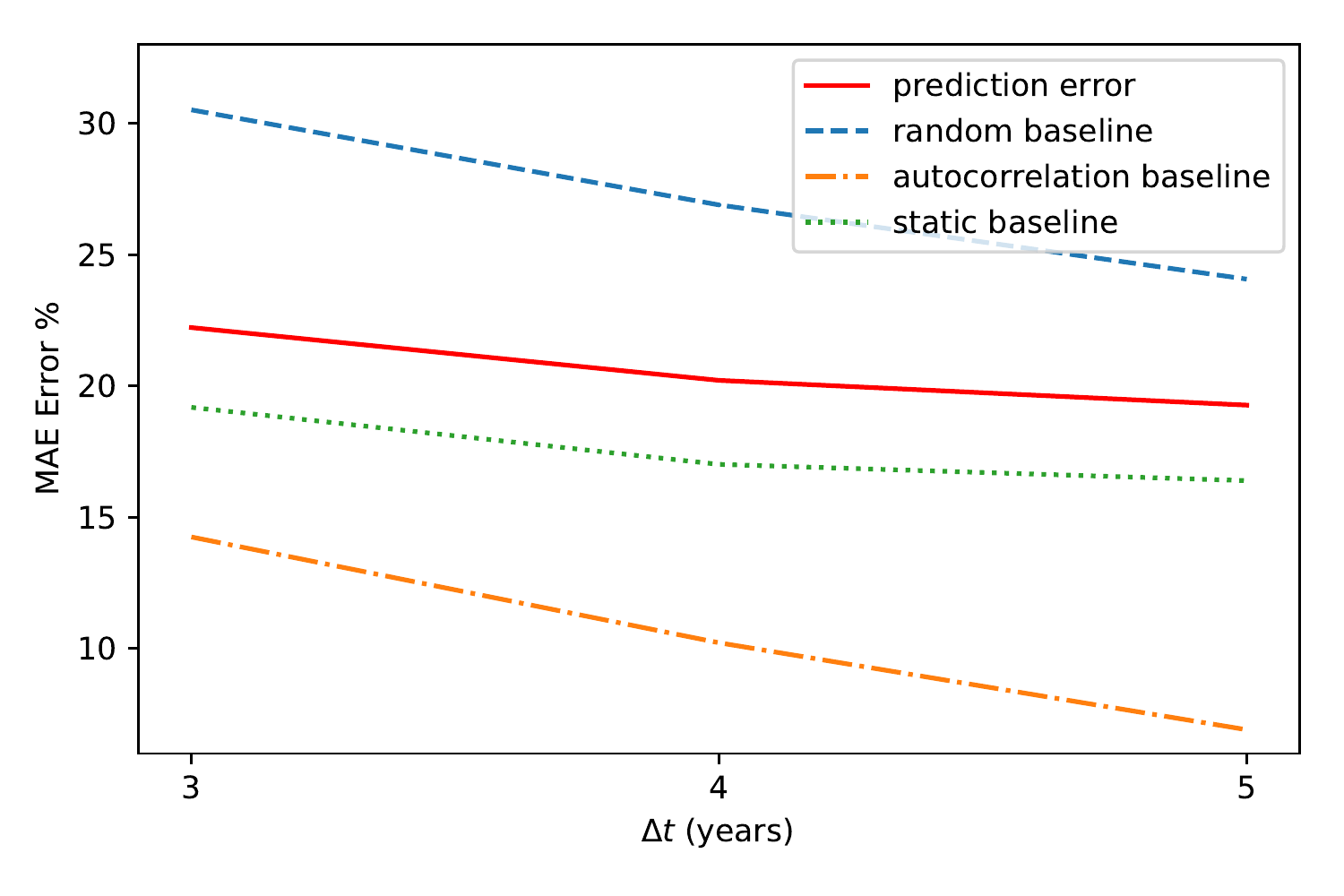}}
    \subfloat[\noreg{4} logPRODY predictions]{
        \includegraphics[clip, trim=11 16 14 14, width=\scalereg]{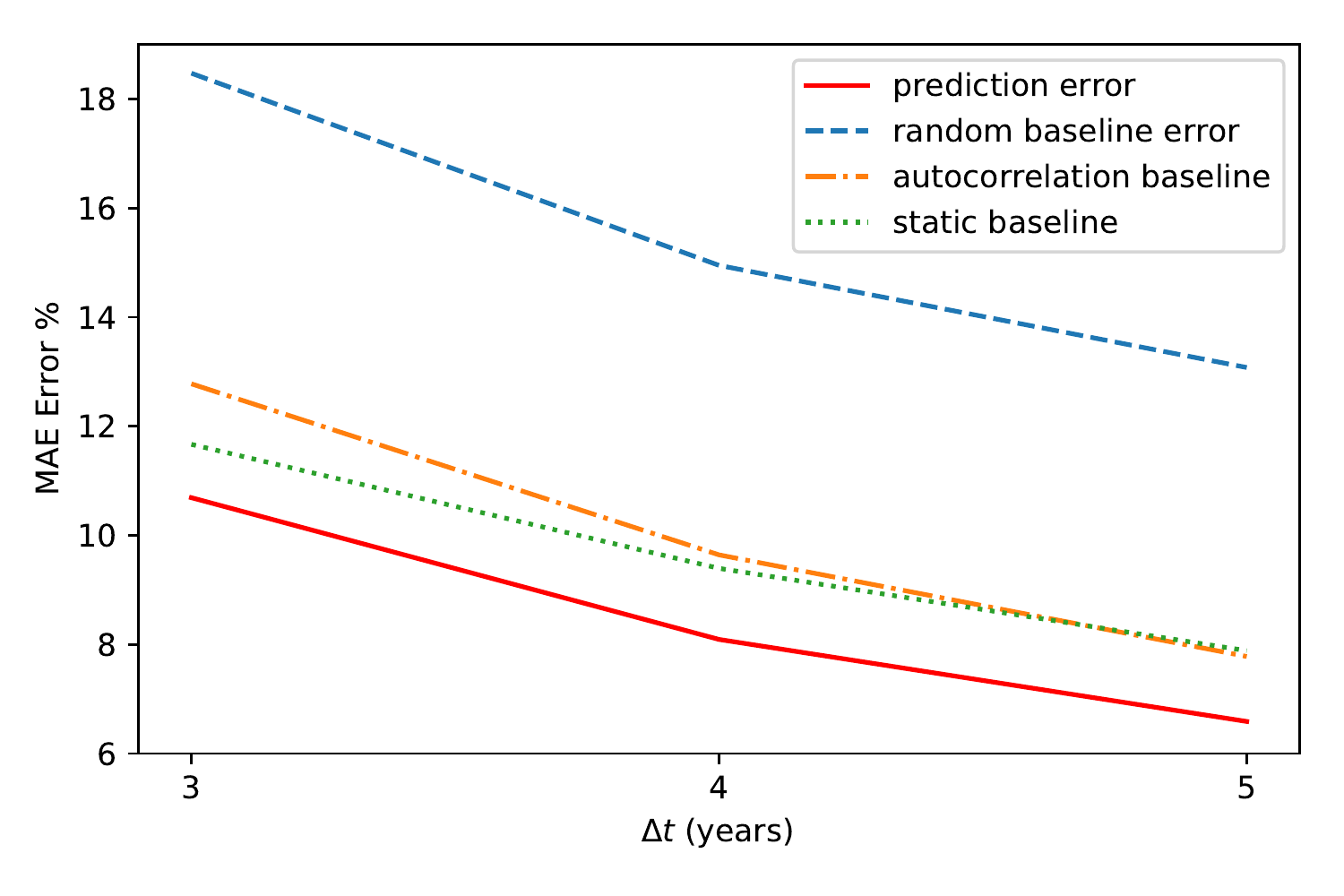}}
    \hspace{0mm}
    
    \subfloat[\noreg{6} Complexity predictions]{
        \includegraphics[clip, trim=11 16 14 14, width=\scalereg]{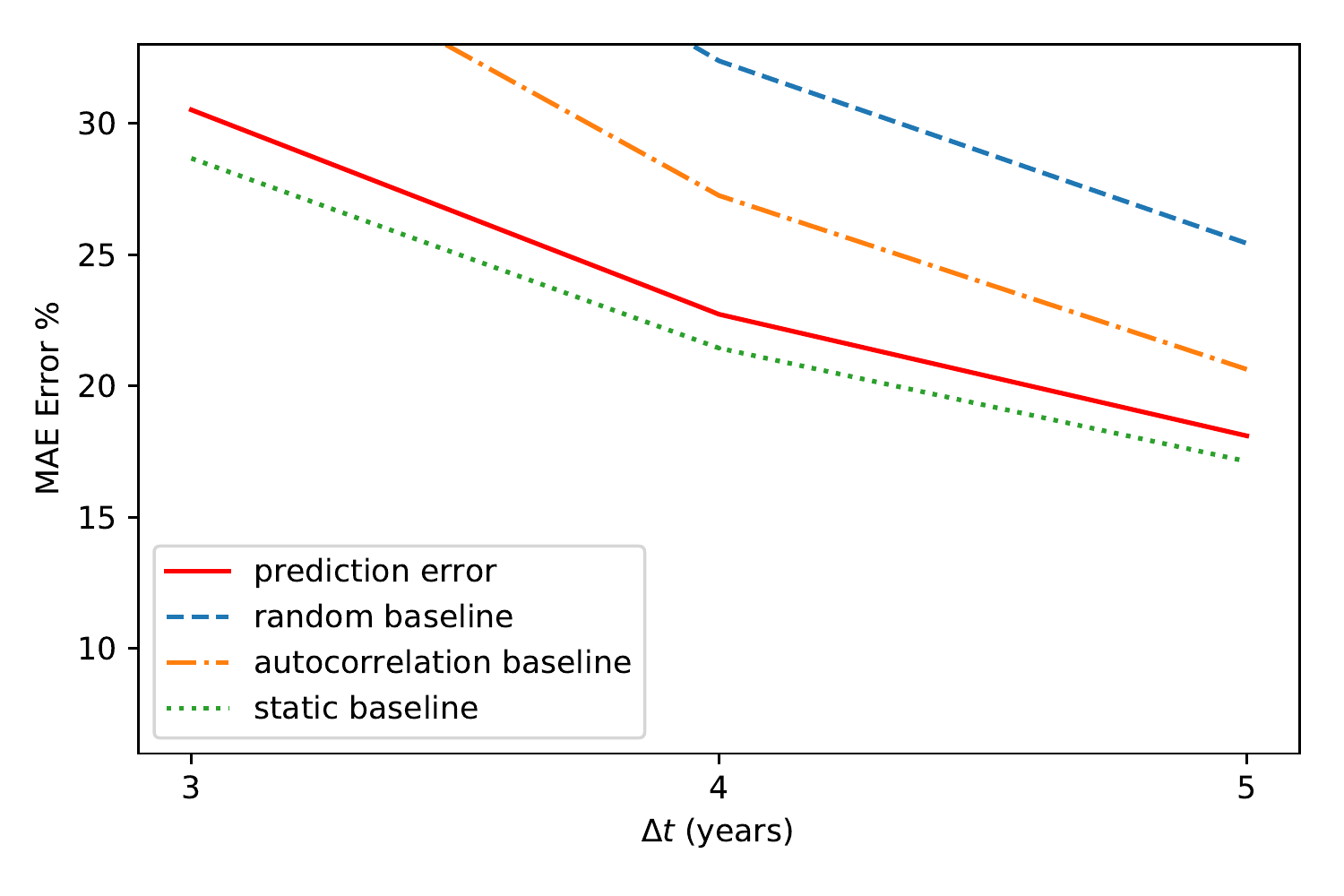}}
    \subfloat[\noreg{6} logPRODY predictions]{
        \includegraphics[clip, trim=11 16 14 14, width=\scalereg]{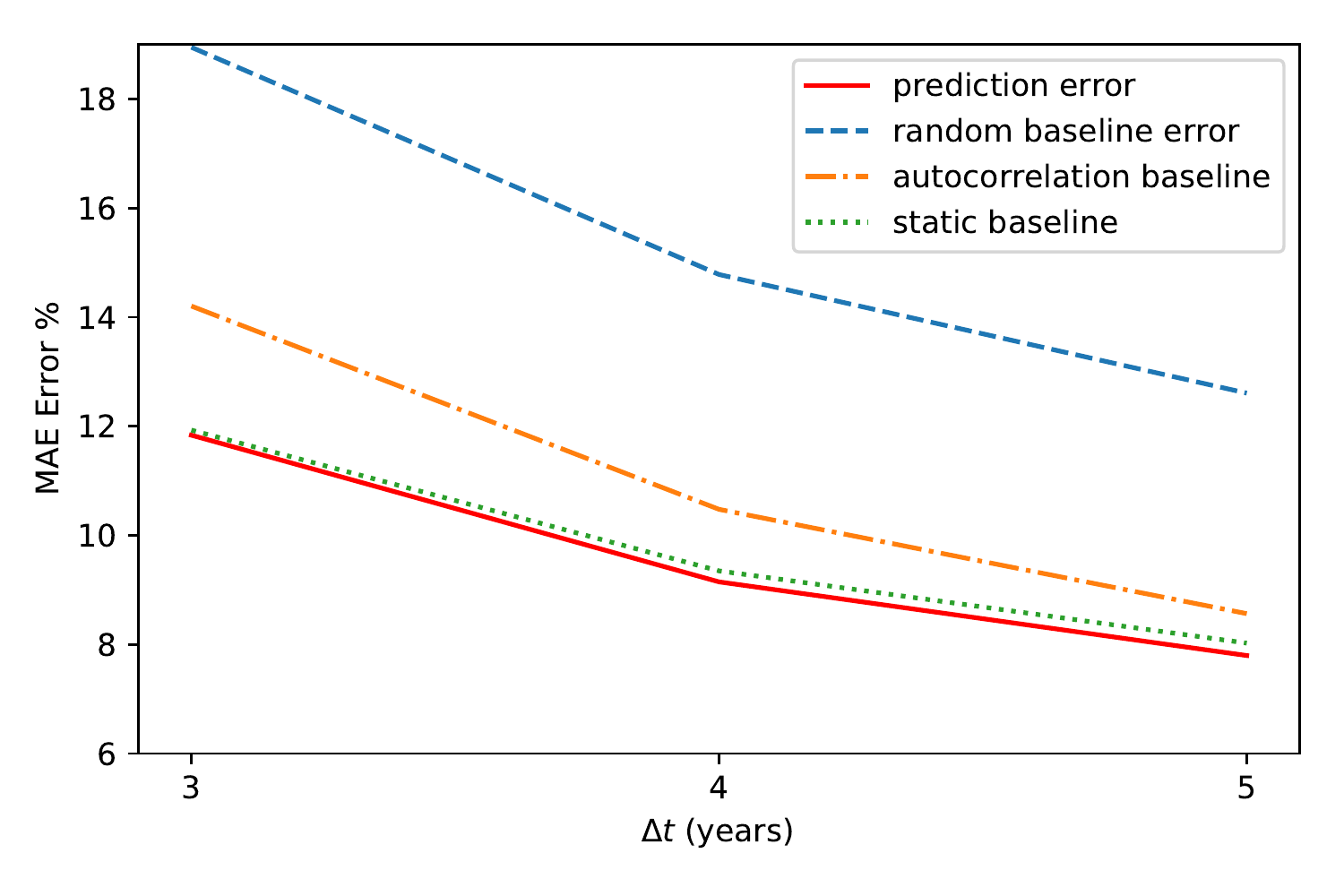}}
    \hspace{0mm}
    
    \subfloat[\reg{6} Complexity predictions]{
        \includegraphics[clip, trim=11 16 14 14, width=\scalereg]{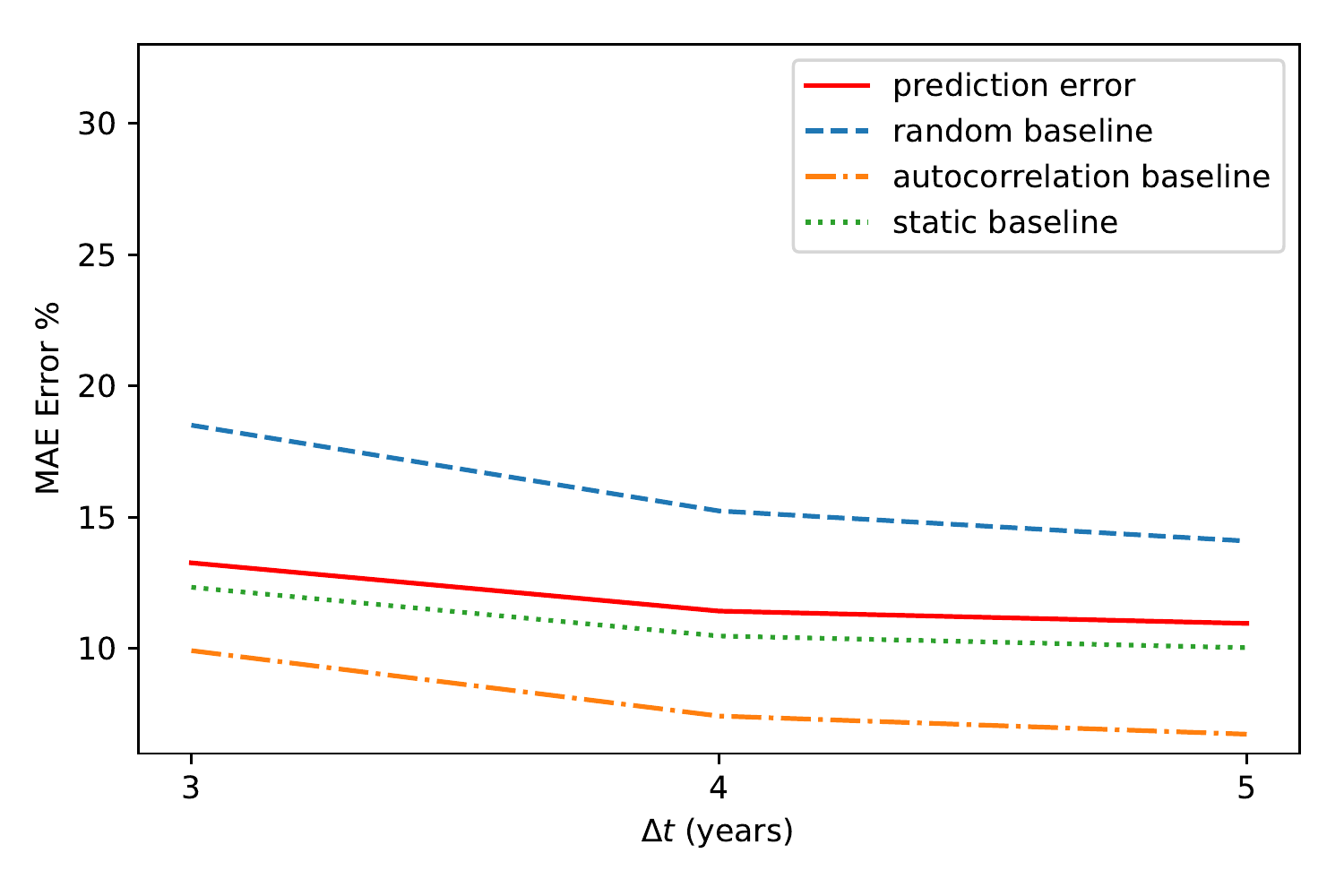}}
    \subfloat[\reg{6} logPRODY predictions]{
        \includegraphics[clip, trim=11 16 14 14, width=\scalereg]{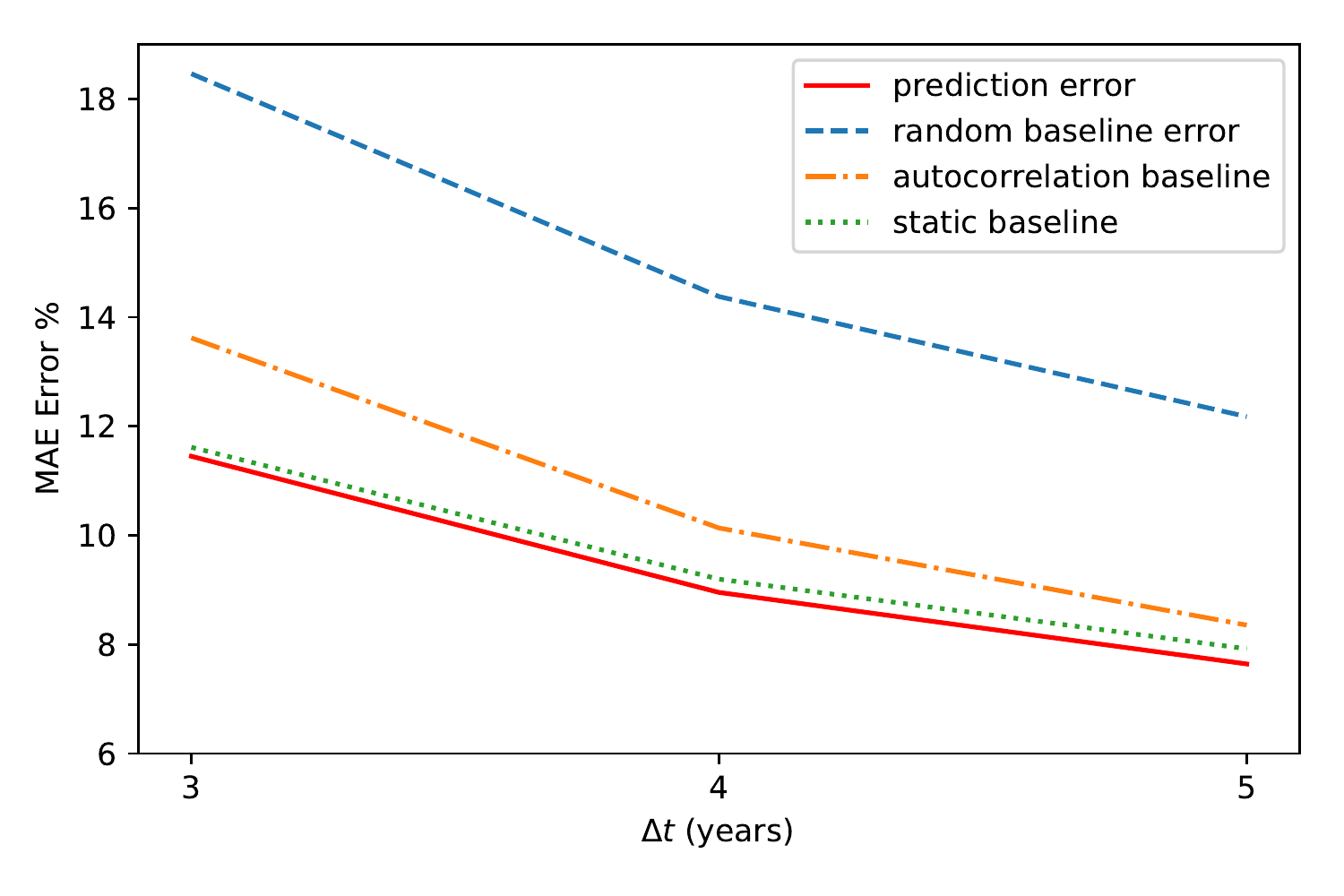}}
    \hspace{0mm}
    \caption{\spsb\ predictions on products. We predicted future values of \logprody\ and Complexity on the log(Complexity)-logPRODY plane (\emph{not} using ranking) with backtested \spsb\ at $\Delta t = 3,4,5$ years in the future. We used the three datasets \reg{6},\noreg{6} and \noreg{4}. On the vertical axis, the Mean Average Error of the prediction (MAE). Three baselines are shown. The first one, called \qt{random}, consists of predicting displacement by randomly selecting one available analogue. The second, called \qt{autocorrelation}, consists of predicting the next displacement of a product to be exactly the same as the last observed one. The last, called \qt{static} predicts 0 displacement for every product.
    \paneldeflong{a,c,e}{left} Complexity predictions are always worse than both the static baselines, and worse than the autocorrelation one in \reg{6} and \noreg{4}. This might signify that observed changes in Complexity mostly caused by random noise. Very interesting is the good result of the autocorrelation baseline: this suggests that Complexity changes over time might be autocorrelated. Finally, prediction accuracy is significantly better for regularized data. It can be interpreted as a signal that, by reducing the noise, the motion becomes more predictable.
    \paneldeflong{b,d,f}{right} \logprody\ predictions are significantly better than random predictions in all cases. Predictions are significantly better than all baselines for \noreg{4}, and slightly but systematically better than the static prediction for the other two datasets. We interpret this as a clue that \logprody\ change over time actually signals a change in market structure, as discussed in \ref{subsection:predictions}. These results also confirm that the \logprody\ model performs significantly better on \noreg{4}.
}
    \label{fig:predictions:predictions}
\end{figure}

\setlength{\scalereg}{.8\textwidth}
\begin{figure}[H]
\centering
    \centering
        \includegraphics[width=1\scalereg]{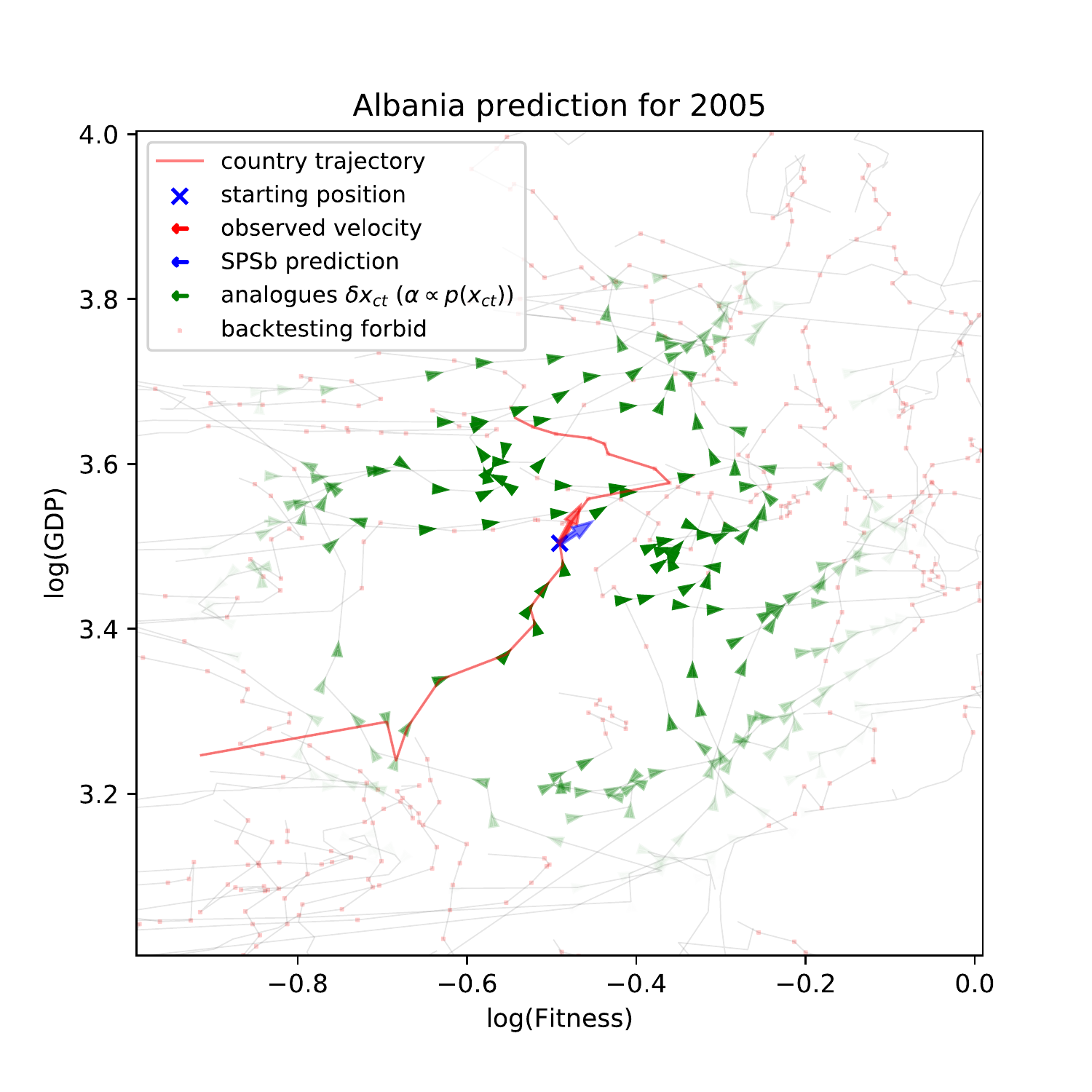}
\caption{An example of \spsb\ prediction. A crop of the Fitness-\gdp\ plane is shown; in light grey the trajectories of countries on it. In red, the trajectory of the country under examination, in this case, Albania. An x marks the position of Albania at time $\thist$ of prediction, 2005. The prediction is the average of all the available analogues, i.e. the observed trajectories of countries at times $t^{\text{past}} < \thist$. The analogues are represented in green (not to scale), and the opacity is proportional to their weight in the final prediction. Analogues excluded from the calculation because are observed in the at times $t^{\text{future}} \geq \thist$ are represented as red dots. A blue arrow represents the predicted displacement on the plane (for both \gdp\ and Fitness), while a red arrow represents the observed displacement during $\Delta t$.
}
\label{fig:predictions:methods}
\end{figure}

\section{Materials and Methods}
\subsection{Fitness and Complexity algorithm}
\label{subsection:fitcomp_def}
As discussed in \refsection{subsection:introduction}, Fitness and Complexity measures are calculated from the $\mcp$. This matrix is intended to be binary, with $\mcp=1$ if country $c$ is an exporter of product $p$, and 0 elsewhere. To measure how significant the exports of $p$ are for a given country, literature turns to the $\rca_{cp}$, where the acronym stands for \define{Revealed Comparative Advantage}, or Balassa index \cite{Balassa1965}, and we defined the weighs. If we define the value in dollars of product $p$ exported by country $c$ as $\exm$ (also known as the \define{export matrix}), then the Balassa index is defined as:
\begin{equation}
 \rca_{cp} = \frac{\frac{\exm_{cp}} {\sum_j \exm_{cj}}} {\frac{\sum_i \exm_{ip}} {\sum_{kl} \exm_{kl}}}.
 \label{eq:rca_definition}
\end{equation}
We take the ratio between the exports of $p$ done by country $c$ and total exports of $c$, and divide it by the world-average of this same ratio. Traditionally, the thresholding of this matrix returns the $\mcp$:
\begin{equation}
\mcp = 
     \begin{cases}
       1 &\quad\text{if}\quad\rca\cp \geq 1,\\
       0 &\quad\text{otherwise.}\\ 
     \end{cases}
     \label{eq:mcp_definition}
\end{equation}
This is the definition we refer to when mentioning \define{unregularized} data. Because both $\exm$ and $\rca$ are noisy matrices, a new procedure procedure for deriving a regularized $\mcp$ has been introduced, as explained in \refsection{subsection:hmm}.\newline
We mention in \refsection{subsection:introduction} that the $\mcp$ matrix is nested, and this observation is crucial to the definition of the Fitness-Complexity Algorithm because of two important implications. The first one is that observing a $p$ being exported by a very diversified country $c$ is uninformative, while if $c$ is poorly diversified we have good reason to think that the product should be a low-Complexity one. On the other hand, if $p$ is only exported by high-Fitness countries, chances are that it should be assigned high Complexity. The algorithm itself is a map that is iterated to convergence on the $\mcp$, and it embeds the former considerations with a non-linearity. The equations of the map are:
\begin{align}
F_c^{(0)}& =1 \quad \forall c, &C_p^{(0)}&=1  \forall p. \\
\tilde{F}_c^{(n)}& =\sum_p \mcp C_p^{(n-1)}, &\tilde{C}_p^{(n)}&=\frac{1}{\sum_c \mcp \frac{1}{F_c^{(n-1)}}}\\
F_c^{(n)}& =\frac{\tilde{F}_c^{(n)}}{\langle \tilde{F}_c^{(n)} \rangle_c},  &Cp^{(n)}&=\frac{\tilde{C}_p^{(n)}}{\langle \tilde{C}_p^{(n)} \rangle_p}.
\label{eq:fitcompmap}
\end{align}
Now Fitness of country $c$ is defined as the plain sum of Complexities of products exported by $c$. Complexity of product $p$ is instead bound by the equations to be less than the lower Fitness found among the exporters of $p$. Additionally, the more exporters of $p$, the less its Complexity. Convergence of the map can be defined numerically in various ways \cite{Pugliese2016,Wu2016}, and the stability of the metric with respect to noise has been studied in \cite{Battiston2014,Mariani2015}.

\subsection{\Logprody}
\label{subsection:logprody_def}
\Logprody\ is a modification of the PRODY index proposed by Hausmann \cite{Hausmann2007}, who employed it to investigate the relationship between exports and growth of a country. logPRODY is defined, for a product $p$, as follows:
\begin{equation}
\lp_p \equiv \sum_c \frac{\rca_{cp} \log_{10}(\gdp_c)}{\sum_j \rca_{jp}} = \sum_c \nrca_{cp} \log_{10}(\gdp_c),
\label{eq:logprodydef}
\end{equation}
where \rca\ is the Balassa index explained in \refsection{subsection:fitcomp_def}, \refeq{eq:rca_definition}.
The Hausmann's PRODY is defined the same way, except that $\log_{10}(\gdp_c)$ is replaced by $\gdp_c$ in the sum. We employ logarithms because the numerical distribution of \gdp s spans several orders of magnitude, and a geometric average contributes to the stability of the measure \cite{Angelini2017}. Note that we defined $\nrca_{cp} = \rca\cp \ \sum_j \rca_{jp}$, the \define{normalized \rca}. Comparing this quantity with the definition of \rca, we can see that normalization removes the effect of numerator from \refeq{eq:rca_definition}. In other words, $\nrca\cp$ is proportional to the ratio between the exports of $p$ done by country $c$ and total exports of $c$. The more product $p$ contributes to total exports of $c$, the more $c$ will be weighed in \logprody$_p$. Further considerations about this measure can be found in \cite{Angelini2017}.

\subsection{\clplong\ motion model}
\label{subsection:clpmotion}
Products can be represented as points on the \clplong\ (\clp) plane. Their aggregate motion in time, averaged as a vector field $\vfield$ can be seen in \reffigure[s]{fig:fields_gradients}\refpanel{a,c,e}. In those figures, the \clp\ plane has been divided into a grid of cells, and we averaged the displacement vector of all products for each cell\footnote{Note that all axes in \reffig{fig:fields_gradients} are labeled as rank$(\cdot)$. This is because Complexity and \logprody\ can be badly behaved, and the standard treatment is to use tied ranking, instead of the observed value, when calculating this model.}. This motion can be modeled with a potential-like equation \cite{Angelini2017}. One first needs to define the Herfindahl index \cite{WAKelly1981}:
\begin{equation}
H_p = \sum_c \left( s_{cp} \right)^2 ; \qquad s_{cp} = \frac{\exm_{cp}}{\sum_c \exm_{cp}}
\label{eq:herfindahl_def}
\end{equation}
where $\exm\cp$ is the export matrix, defined in \refsection{subsection:fitcomp_def}. The Herfindahl index measures the competitiveness of a market by summing the square of the market shares of each participant to the market. It ranges from 1 (for a monopoly) to $1/N$ (the case of $N$ participants all with equal market share). When defined as in \refeq{eq:herfindahl_def}, it refers to the total market share of countries. Averaging the Herfindahl index per cell on the \clp\ plane produces a scalar field, $H$, for which one can compute the gradient with respect to the $C$ (Complexity) and $\lp$ (\logprody) coordinates on the plane. Then the model explaining $\vfield$ is:
\begin{equation}
  \vec{v} \simeq - k_{C} \frac{\partial H}{\partial C} \vec{C} - k_{\lp} \frac{\partial H}{\partial \lp} \vec{\lp} \equiv -\vec{\nabla}_{k} H \label{eq:hfield-grad-def}
\end{equation}
where $k_{C}$,$k_{\lp}$ are two scalar constants. This implies that the average velocity of products $\vfield$ points towards area of lower $H$, i.e. higher competition on the \clp\ plane. The lines in \reffigure{fig:fields_gradients} show respectively where $\vfield$ is minimum and where $H$ is minimum for each column of the grid.\newline

The interpretation given to this model in \cite{Angelini2017} is that \logprody$_p$ serves as a proxy for the global \define{market structure} of product $p$. The full market structure is defined by the distribution of the weights of \logprody$_p$ across countries. As mentioned in \refsection{subsection:logprody_def}, these weights are given by the \nrca$\cp$ and they are proportional to the competitive advantage of country $c$ in making product $p$. The market structure that maximizes $H$, or competition, is named \define{asymptotic} in \cite{Angelini2017}, and it depends on Complexity. Low-Complexity products typically show an asymptotic distribution of comparative advantage that is uniform across all countries, or sometimes mildly peaked on low-Fitness countries. High-Complexity products show instead a sharp peak of comparative advantage on high-Fitness countries. The name asymptotic comes from the observation that whenever the market structure of a product is different from the asymptotic, it tends to revert to it. In doing so, it increases competition ($H$). \Logprody\ is by definition the expectation value of the \gdp\ on the distribution of comparative advantage, so its value tends to revert to the value it assumes on the asymptotic distribution.\newline
Interpretation for the horizontal displacements (along the Complexity axis) is, instead, less clear-cut. This difference in interpretability between \logprody\ and Complexity displacements plays a role into our discussion of \refsection{subsection:predictions}.

\subsection{\spsb}
\label{subsection:spsb}
As mentioned in \ref{subsection:convergence}, \define{Bootstrapped Selective Predictability Scheme} (\spsb) is a prediction technique allowing quantitative forecast of \gdp\ growth for a country by averaging the growth of countries nearby on the \fglong\ (\fg) plane \cite{Tacchella2017,Tacchella2018}. We will describe the algorithm in detail here. Given $\fgpos_{\thisct}$, the position of country $\thisc$ in the \fg\ plane at time $\thist$, we want to forecast $\delta \analog_{\thisct}$ \footnote{Note that while the position on the \fg\ plane is vectorial ($\fgpos$), we are referring to the displacement as a scalar ($\delta \analog$). This is because we want to keep the formalism of the original work, which is concerned only with displacement along the \gdp\ direction. Nothing forbids to forecast displacement along any arbitrary direction, though. In that case, the displacement would have to be a vector quantity.}, the future displacement of country $\tilde{c}$ from time $\thist$ to $\thist + \Delta t$. To do so, we consider the set of observed past observations $(\delta \analog_{c,t}, \fgpos_{c,t})$ on the \fg\ plane, which we will call \define{analogues}. Note that, if one wants to rigorously implement a backtesting procedure, only the analogues for which $t < \thist$ are allowed. It is possible to bootstrap an empirical probability distribution for $\delta \analog_{\thisct}$ in two steps:

\begin{enumerate}
\item Sample with repetition the $N$ available analogues with a probability distribution $p$ given by a gaussian kernel
centered in $\analog_{\thisct}$, i.e. the probability of sampling the analogue displacement $\delta \analog_{c,t}$ is:
\begin{align}
&p(\delta \analog_{c,t} | x_{c,t}) = \mathcal{N}(\fgpos_{\thisct} - \fgpos_{c,t} | 0, \sigma),\\
&\mathcal{N}(\vec{z}| \vec{\mu}, \sigma) = \frac{1}{\sigma \sqrt{2\pi}}\text{exp}\left(\frac{(\vec{z} - \vec{\mu})^2}{2\sigma^2}\right).
\end{align}
Note that the probability of sampling depends only on the Euclidean distance between $\fgpos_{\thisct}$ and the position of the analogue.

\item Sample $B=1000$ bootstraps with the above procedure (bootstrap) and average the displacements per bootstrap. The global distribution of these
averages is the empirical probability distribution for $\delta \fgpos_{\thisct}$. The mean of the distribution is used as the prediction value and the standard deviation as the uncertainty on the forecast.
\end{enumerate}

\subsection{\nwkdelong}
\label{subsection:nwkde}
\nwkdelong\ was originally introduced in 1964 \cite{Nadaraya1964,Watson1964}. Its purpose is to estimate the conditional expectation of a variable $Y$ relative to a variable $X$, which we will denote as $E(Y | X)$, in the hypothesis that the probability distributions $f(X,Y)$ and $f(X)$ exist. If one has $n$ sampled observations $(X_1,Y_1), \ldots, (X_n,Y_n)$ (where $X$ can be multivariate), the regression model is:
\begin{equation}
    Y_i = m(X_i) + \epsilon_i
    \label{eq:nwkr_model}
\end{equation}
where $m(x)$ is a (yet) unknown function and the errors satisfy these hypotheses:
\begin{equation}
    E(\epsilon)=0; \qquad \text{Var}(\epsilon)=\sigma^2_\epsilon; \qquad \text{Cov}(\epsilon_i,\epsilon_j)=0 \quad \forall i \neq j.
\end{equation}
One can try to approximate the probability distributions with a kernel density estimation:
\begin{align}
    f(X,Y) &\approx \hat{f}(X,Y) = \frac{1}{n} \sum_{i=1}^n K_h (X-X_i) K_h(Y-Y_i) , \label{eq:nwkde_approx}\\
    f(X) &\approx \hat{f}(X) = \frac{1}{n} \sum_{i=1}^n K_h (X-X_i). \label{eq:nwkde_approx2}
\end{align}
where $K_h(x) = K(x/h)/h$ is a \define{kernel}, i.e. a non-negative function such that $\int K(x) dx = 1$, and $h>0$ is called \define{bandwidth} and scales the kernel to provide smoothing to the regression. In this paper we will use only one type of kernel, the \define{gaussian} (also known as \define{radial basis function}): $K(x)=e^{-x^2}$.
The conditional expected value can therefore be approximated, using \refeq{eq:nwkde_approx},\ref{eq:nwkde_approx2} as:
\begin{align}
    E(Y | X) &= \int Y f(Y|X) dY = \int y\frac{f(X,Y)}{f(X)} dY \\
    &\approx \int \frac{Y \sum_{i=1}^n K_h (X-X_i) K_h(Y-Y_i)}{\sum_{i=1}^n K_h (X-X_i)} dY \\
    &= \frac{\sum_i K_h (X-X_i) \int Y K_h(Y-Y_i) dY}{\sum_i K_h (X-X_i)} \\
    &= \frac{\sum_i K_h (X-X_i) Y_i}{\sum_i K_h (X-X_i)} \equiv \hat{E}(Y | X). \\
\end{align}
Therefore we can rewrite $m$ in \refeq{eq:nwkr_model} as:
\begin{equation}
m_h(x)=\frac{\sum_i K(\frac{x-X_i}{h}) Y_i}{\sum_i K(\frac{x-X_i}{h})}.
\end{equation}

\subsection{\hmm\ regularization}
\label{subsection:hmm}
As explained in \refsection{subsection:fitcomp_def}, the traditional way to calculate the $\mcp$ matrix consists of calculating the \rca (\refeq{eq:rca_definition}) and then thresholding it (\refeq{eq:mcp_definition}). This procedure introduces noise in the matrix because very often the value of \rca\ fluctuates around the threshold. By introducing time in the estimation of the $\mcp$ it is possible to mitigate this problem. The procedure has been introduced in \cite{Tacchella2018}, and it consists of modelling each \rca$\cp$ time series as the emission probabilities of hidden states in a Hidden Markov Model \cite{rabiner1986introduction} (\hmm). The competitive advantage of a given country $c$ in making product $p$ is represented as a series of 4 quantized \qt{developement stages}, obtained by calculating the quantiles of the \rca$\cp$ time-series. We will call this quantized matrix \rca$^q$ To each of these development stages corresponds a probability to express a given value of \rca$\cp$. Countries transition between these development states with a Markov process that has transition matrix $T$. Both $T$ and the parameters of the \rca\ distribution are estimated with the Baun-Welch algorithm \cite{rabiner1986introduction}. Additionally, one separate model is evaluated for each country. The algorithm produces one \rca$\cp^q$ matrix for each year of observation, containing the most probable development stage at each timestep. The matrices can then be binarized. It can be shown that this regularization technique reduces noise and increases the predictive performance of the \spsb\  algorithm \cite{Tacchella2018}.


\subsection{Datasets and product digits}
\label{subsection:datasets}
In this work, we use a dataset containing all the information of the $\exm$ matrix (from which all the Economic Complexity metrics can be calculated). We call it BACI, and it is documented in \cite{Gaulier2010}. The original data in BACI comes from UN-COMTRADE, and it has been further elaborated by CEPII, which sells the right to use it. The elaborated version of the dataset is not in the public domain, but a free version without data cleaning is available on the BACI section of the organization's website \cite{datasetBACI}. 149 countries are included in our analysis, spanning 21 years from 1995 to 2015. Products are classified by UN-COMTRADE according to the Harmonized System 2007 \cite{datasetWorldCustomsOrg} (HS2007). HS2007 is divided in 16 Sections, which are broad categories such as, e.g., \qt{Vegetable Products}, \qt{Textiles}, \qt{Metals}, and so on. Products are then hierarchically denoted each by a set of 6-digit codes. The code is divided into three 2-digit parts, each specifying one level of the hierarchy: so the first part (Chapter) indicates the broadest categories, such as e.g. \qt{Cereals} (10xxxx). The second two digits (Heading) specify further distinctions in each category, for example, \qt{Rice} (1006xx). The last two digits (Subheading) are more specific, e.g. \qt{Semi-milled or wholly milled rice, whether or not polished or glazed} (100630). For the analysis mentioned in the paper, we look at data for products aggregated at both 4-digit level (1131 products retained) and 6-digit level (4227 products).  Data cleaning procedures outside of the \hmm\ regularization mentioned above consist in the elimination of extremely small countries and countries with fragmented data; aggregation of some product categories that are closely related, and (for what we call non-regularized data) a very simple regularization of the $\mcp$ matrices based the recognition and substitution of fixed handmade patterns. GDPpc data has been downloaded from the World Bank Open Data website \cite{datasetWorldBank}.

\section{Conclusions}
\label{subsection:conclusions}
In this work, we focused on the analysis of Product Complexity, which had received little attention since \cite{Angelini2017}. The application of the motion model to the 6-digit data set with and without \hmm\ regularisation seems to indicate that much of the change in Complexity over time is due to noise. Further analysis will be certainly needed on this topic, as it could lead to a better understanding of the Complexity measure as discussed in \ref{subsection:predictions}. We suggest that these results should be strengthened and confirmed in future work by an evaluation of the quantity of noise might be carried out, in the fashion of \cite{Battiston2014,Tumminello2011}. Insights gathered this way might be used to calibrate a model that evaluates the effect of noise on Complexity change over time. Also very interesting is the finding that changes in Complexity might be autocorrelated over time. Further analysis is needed to clarify whether this is true, and if appropriate to understand the causes of the autocorrelation.\newline
Applying \spsb\ to the \clp\ plane seems to confirm the findings of \cite{Angelini2017} regarding the meaning of \logprody\, and gives further grounds to argue that changes in Complexity over time are not relevant. The same suggestions as before apply: further validation with a study of the noise is probably a good research path. We analysed the change in nestedness caused by the \hmm\ regularisation technique on the $\mcp$ matrices, and thoroughly validated the statistical significance of the difference with several null models. We suggest that aggregating data from 6 to 4-digit level might have a regularising effect.\newline
Finally, in order to be able to apply \spsb\ to a data set larger by one order of magnitude than what was previously done, we developed proof that \spsb\ itself converges, for a high number of iterations, to a well-known statistical learning technique, \nwkde. The two techniques can be used interchangeably. \nwkde\ has the advantage of being significantly faster, and of producing a deterministic result. The proof also has the benefit of further clarifying the nature of \spsb. This technique belongs to the same family of algorithms that predict by similarity based on distance, such as \nwkde\ and k-nearest neighbours. We suggest that regression trees might do well in its place, too. We also suggest that a further technical development in this field might be the introduction of one of the many flavours of variable-bandwidth \nwkde\ techniques because of the significant changes in density of analogues over the considered data sets.

\vspace{6pt} 

\subsection*{Supplementary information}
We released custom code for \nwkde\ calculations \cite{ectools} and a wrapper that allows using FALCON\cite{Beckett2014} in Python 3.6\cite{pyfalcon}. Data availability is discussed in \refsection{subsection:datasets}.

\subsection*{Acknowledgements}
We would like to thank Professor Luciano Pietronero, Dr Andrea Gabrielli, Dr Andrea Zaccaria, Dr Andrea Tacchella, Dr Dario Mazzilli, for enlightening conversations and for freely exchanging their findings and data with us. Thanks to Anshul Verma for the many useful discussions. We also wish to thank the ESRC Network Plus project “Rebuilding macroeconomics”.

\subsection*{Abbreviations}
The following abbreviations are used in this manuscript:\\
\noindent \gdp: Gross Domestic Product\\
\spsb: Bootstrapped Selective Predictability Scheme\\
\hmm: Hidden Markov Model\\
\nwkde: \nwkdelong\\
$\lp$: \logprody\\
$C$: Complexity\\
$\mcp$: export bipartite network adjacency matrix\\
\fg: Fitness-\gdp\\
\clp: Complexity-\logprody\\
\rca: Revealed Comparative Advantage\\
\nrca: Normalized \rca\\
\exm: EXport Matrix\\

\appendix
\section{Country predictions via the products}
\label{appendix:country_predictions}
Even though a definitive interpretation for both Complexity and \logprody\ is lacking, if predictions on the trajectories of products are better than random, one can try and use them to make predictions on the countries' trajectories. By definition, a country's Fitness is equal to the sum of the complexities of its exports (see \refsection{subsection:fitcomp_def}), i.e.
\begin{equation}
 F_c = \sum_p \mcp Q_p,
 \label{eq:predictions:fit}
\end{equation}
while countries' \gdp's are connected to the \logprody s via
\begin{equation}
 \logprody_p = \sum_c \frac{\rca_{cp} log_{10}(\gdp_c)}{\sum_j \rca_{jp}} \equiv \sum_c \nrca_{cp} log_{10}(\gdp_c),
 \label{eq:predictions:gdp}
\end{equation}
where we defined $\nrca_{cp} \equiv \rca_{cp} / \sum_j \rca_{jp}$. Therefore, if we can find $\nrca^{-1}$ such that $\nrca^{-1}\nrca = 1$, we can invert the relation and obtain:
\begin{equation}
 log_{10}(Y_c) = \sum_p \nrca^{-1}_{pc} \logprody_p.
\end{equation}
We can then feed our estimates of future positions of products to these equations, to obtain an estimate on future positions of countries on the \fg\ plane. Because of the lack of predictive power described in \refsection{subsection:predictions}, country predictions are worse than all baselines (result not shown in this work). Furthermore, it is known in general from the statistical learning literature\cite{friedman2001elements}, and in particular for Economic complexity\cite{Tacchella2017} that averaging the prediction of two different models can improve significantly the error of a regression. Averaging our countries' predictions with the predictions made by \spsb\ on the \fg\ plane results in worse performance, thus we argue that the product's predictions are tainted by large amounts of noise. This noise comes primarily from the locally disorderly motion in the \clp\ plane, but there is another important source of noise too. An important contribution to the change in Fitness is due to new products being exported (or lost) over time. But in a backtesting, the $\mcp$ and $\nrca$ matrices fed to \refequation{eq:predictions:fit} contain only information about products exported at the initial time. This is true for the \gdp\ too, if one substitutes the $\mcp$ matrix with the $\nrca$ in \refeq{eq:predictions:gdp}.

\bibliographystyle{plain}

\bibliography{biblist.bib}

\end{document}